\documentclass[11pt]{article}
\usepackage{amsfonts}
\usepackage{amssymb}
\usepackage{amsmath}
\usepackage{amsthm}
\usepackage{epsfig}

\newlength{\bredde}
\def\slash#1{\settowidth{\bredde}{$#1$}\ifmmode\,\raisebox{.15ex}{/}
\hspace*{-\bredde} #1\else$\,\raisebox{.15ex}{/}\hspace*{-\bredde} #1$\fi}
\newcommand{\be}{\begin{equation}}
\newcommand{\ee}{\end{equation}}
\newcommand{\bea}{\begin{eqnarray}}
\newcommand{\eea}{\end{eqnarray}}
\newcommand{\nn}{\nonumber}
\newcommand{\Dirac}{\rlap{\hspace{-1.2mm} \slash} D}
\newcommand{\eins}{\leavevmode\hbox{\small1\kern-3.8pt\normalsize1}}
\newcommand{\e}{\mbox{e}}
\newcommand{\erf}{\mbox{erf}}
\newcommand{\sign}{\mbox{sign}}
\newcommand{\Om}{\Omega}
\newcommand{\la}{\lambda}

\newcommand{\hm}{\hat{m}}
\newcommand{\hz}{\hat{z}}
\newcommand{\hx}{\hat{x}}
\newcommand{\hy}{\hat{y}}
\newcommand{\ha}{\hat{a}}

\newcommand{\eps}{\epsilon}
\newcommand{\sect}[1]{\setcounter{equation}{0}\section{#1}}

\def\Tr{{\mbox{Tr}}}

\def\Pf{\mbox{Pf}}

\begin{document}
\topmargin -1.4cm
\oddsidemargin -0.8cm
\evensidemargin -0.8cm
\title{\Large{{\bf
Random Matrix Theory for the Hermitian Wilson Dirac Operator\\
and the chGUE -- GUE Transition
}}}

\vspace{1.5cm}

\author{~\\{\sc Gernot Akemann}$^{1}$ and {\sc Taro Nagao}$^{2}$
\\~\\
$^1$Department of Physics,
Bielefeld University,
Postfach 100131,
D-33501 Bielefeld, Germany\\~\\
$^2$Graduate School of Mathematics, Nagoya~University,
Chikusa-ku, Nagoya~464-8602, Japan
}
\date{}
\maketitle
\vfill
\begin{abstract}
We introduce a random two-matrix model interpolating between a chiral Hermitian
$(2n+\nu)\times (2n+\nu)$ matrix and a second Hermitian matrix without
symmetries. These are taken from the 
chiral Gaussian Unitary Ensemble (chGUE) and Gaussian Unitary
Ensemble (GUE), respectively. In the microscopic large-$n$ limit in
the vicinity of the chGUE (which we denote by weakly non-chiral limit) this
theory is in one to one correspondence to the partition function of
Wilson chiral perturbation theory in the epsilon regime, such as the
related two matrix-model previously introduced in \cite{DSV,ADSV}. 
For a generic number of flavours and rectangular block matrices in the
chGUE part we 
derive an eigenvalue representation for the partition function
displaying a Pfaffian 
structure.
In the quenched case with $\nu=0,1$ 
we derive all spectral correlations functions in our model for
finite-$n$, given in terms of skew-orthogonal polynomials. The latter
are expressed as Gaussian integrals over standard Laguerre
polynomials. 
In the weakly non-chiral microscopic limit this yields all
corresponding quenched eigenvalue correlation functions of the
Hermitian Wilson operator.

\end{abstract}
\vfill

\thispagestyle{empty}
\newpage

\renewcommand{\thefootnote}{\arabic{footnote}}
\setcounter{footnote}{0}

\sect{Introduction}\label{intro}

The application of random matrix theory (RMT) to Quantum
Chromodynamics (QCD) first introduced in \cite{SV93} has become much
more sophisticated in the recent past. Starting from a Gaussian RMT
with $N_f$ massless flavours \cite{SV93,JRMT}, the
so-called chiral Gaussian unitary ensemble (chGUE), several
milestones have been taken from which we mention only a few, 
see \cite{JacHB} for recent reviews and more references.
After the universality of the microscopic origin limit 
for non-Gaussian RMT was understood \cite{ADMN},
the computation of unquenched density correlation functions
of the QCD Dirac operator with non-vanishing quark masses followed
\cite{DN97,WGW}, as
well as the computation of individual Dirac eigenvalues
\cite{WGW,DNW,nagaoforr,AD03}. It was
understood \cite{DOTV} that the RMT approach is in one to one correspondence 
to the epsilon regime of chiral perturbation theory (chPT)  \cite{GL87a}
as a limiting theory of low-energy QCD. Corrections to this regime were
computed by including finite volume corrections to the chiral
condensate $\Sigma$ \cite{GL87a}, or to the Pion decay
constant $F$
\cite{ABL} (see \cite{DF} and\cite{LBHW} for recent discussions). 
However, the latter only appears in RMT when extending it to a
two-matrix model, by coupling to a real \cite{AOSV} 
or imaginary chemical potential \cite{DHSS,ADOS}.
Otherwise meson correlation functions have to be
considered in chPT to be sensitive to $F$,
by coupling to the non-zero momentum modes of the Pions. 
Most of the RMT predictions have been checked using Lattice QCD, ranging
from checks of the topology dependence of algorithms
with good chiral properties on small Lattices \cite{Edwards}, up to
fully unquenched two-flavour simulations leading to realistic values
of $\Sigma$ and $F$ \cite{Fukaya},  to where we refer as well as to
\cite{JacHB} for more references.

Very recently a further extension of RMT for QCD has been proposed in
\cite{DSV} in order to include also the effect of finite lattice
spacing $a$ close to the continuum. The corresponding RMT is again a
two-matrix model, as it was discussed in more detail in \cite{ADSV}. 
It is the aim of this paper to investigate this RMT further - in fact
in a slightly modified version - and to show its integrability by
computing all density correlation functions. 

The corresponding framework in continuum effective field theory is
Wilson chiral perturbation theory (WchPT), see \cite{GS} for
introductory lectures and references.
Here, in addition to $\Sigma$ and $F$ three further low energy constants
$W_{6,7,8}$ appear to leading order in the infinite volume limit, that
have to be determined non-perturbatively. 
The spectral properties of the Wilson Dirac operator
were put into focus in \cite{Sharpe2006},
and the epsilon regime in WchPT was first analysed in
\cite{Shindler}. The computation of the quenched spectral
density in the epsilon regime - both 
for the Hermitian Wilson Dirac operator as well as for the
real eigenvalues of the non-Hermitian Wilson Dirac operator -
was performed in \cite{DSV} (see also \cite{SN} for more recent results in
the $p$-regime).  The presentation 
in \cite{DSV,ADSV} was in fact solely based on the WchPT Lagrangian
and corresponding generating resolvents, and not on an RMT computation,
although the two become equivalent in the microscopic limit once the
graded or supersymmetric approach is taken. The extension to the
$N_f=1$ \cite{ADSVLatNf} and to $N_f>1$ \cite{SV2011} 
flavoured spectral densities followed also the
supersymmetric approach. 

Our aim is to partly extend these results 
and to compute all spectral densities
from RMT. The advantage of the approach presented here using
skew-orthogonal polynomials is that an explicit eigenvalue basis can
be  found for the Hermitian Wilson Dirac operator. All spectral correlations
then follow both for finite- and large-$N$, where $N=2n+\nu$ is the
size of the matrices, given the corresponding
skew-orthogonal polynomials can be constructed. We have succeeded in this
program in the quenched approximation and for $\nu=0,1$ as a first step,
where $\nu$ counts the number of exact zero-modes at $a=0$.
We will come back to the open challenges left to our RMT approach in the
conclusions. 

Our investigations presented here have a second motivation. So-called
parametric density correlations have been studied in the RMT
literature for several decades. The reason of interest is that the
classical Wigner-Dyson ensembles, the Gaussian unitary, orthogonal
and symplectic ensemble (GUE, GOE and GSE, respectively) correspond to
Hamiltonian systems without (GUE) or with time-reversal symmetry (GOE
or GSE, depending on the spin). For that
reason the symmetry breaking has been studied 
in terms of a two-matrix model that
interpolates between two classes, see
\cite{PandeyMehta} for the classical papers, or chapter 14 in \cite{Mehta} for a
more comprehensive presentation of the GUE-GOE and GUE-GSE 
transition. 
Several such transitions have been studied
since, including the chiral versions of these transitions \cite{FHN}.

A crucial step is to be able to perform the angular integrations in
the term coupling the two random matrices. This is done using the
Harish-Chandra-Itzykson-Zuber (HCIZ) integral \cite{HCIZ} and
explains, why in all given examples (at least) 
one of the two classes has to possess unitary symmetry.
Once an eigenvalue basis is determined the standard (skew)-orthogonal
polynomial approach can be taken, where in \cite{NF98}
a more general framework for general, non-Gaussian 
weight functions has been developed.

These two-matrix models also called transitive ensembles include
also transitions within the {\it same} symmetry class, that is from
the GUE to GUE or chGUE to chGUE, see e.g. \cite{ADOS} including their
extensions 
with $N_f$ flavours, 
in the context of QCD with imaginary isospin chemical potential in three
and four dimensions. 
As we will show below the RMT corresponding to WchPT in the epsilon
regime in \cite{DSV,ADSV} corresponds (after a minor extension) to the symmetry
transition from the chGUE to the GUE. 
Our paper thus serves also to close a gap within RMT by studying this
transition class. Because of the 
ubiquity of RMT we expect that applications in totally different areas
may follow.

The presentation of the remaining chapters is organised as follows. In
section \ref{MM} we present our two-matrix model, show its relation to
WchPT in the epsilon regime (see also appendix \ref{equiv} for
details) and discuss the relation to the RMT proposed in
\cite{DSV,ADSV}. 
In section \ref{jpdf} we derive the joint probability density function
for the Hermitian Wilson Dirac operator for a general $N_f$-flavour
content and general $\nu\geq0$. 
The solution of our two-matrix model for all density correlation
functions is presented in section \ref{Rk} in terms of skew-orthogonal
polynomials for $N_f=0$ and $\nu=0,1$, see also appendix \ref{SOP-C} for further
details. These are given in terms of Gaussian integral transforms of
standard Laguerre polynomials as they appear in the chGUE
\cite{SV93}. 
The microscopic large-$N$ limit at the origin close to the chGUE is
then presented in 
section \ref{largeN} 
illustrated with several examples, before we present our conclusions
and open problems in section \ref{conc}.

\sect{The Random Matrix Model}\label{MM}

In this section we first introduce our matrix model and explain why it
describes the transition between the different symmetry classes of
random matrices from the chGUE to the GUE. Then we point out the
relation to the related Wilson chiral RMT previously introduced in
\cite{DSV} as well as to WchPT in the epsilon regime.

We want to compute the eigenvalue density correlation functions of
the following Hermitian Wilson Dirac operator:
\be
\Dirac_5 \ =\
\left(\begin{array}{cc}
m\eins_n& W\\
W^\dag&-m\eins_{n+\nu}\\
\end{array}\right)
\ +  \ H\ .
\label{D5Idef}
\ee
Here $W$ is a rectangular matrix of size $n\times(n+\nu)$ with complex
elements, without further symmetry restrictions. The non-negative
parameter $\nu\geq0$ is related to the number of zero-eigenvalues in
the limit to the chGUE. The real parameter $m$ denotes the quark
mass, as will become clearer below. This first part of $\Dirac_5 $
is what we are used to in the chGUE (up to the sign in $m$). The second
part $H=H^\dag$ is a quadratic Hermitian matrix of size $N\equiv 2n+\nu$ with
complex elements. It corresponds to the GUE part of  $\Dirac_5 $,
breaking chiral symmetry.

The matrices $W$ and $H$ are distributed with the following Gaussian
probability measures:
\bea
P_1(W)&=&\exp\left[-\ \frac{1}{2(1-a^2)}\Tr WW^\dag\right]\ ,\nn\\
P_2(H)&=&\exp\left[-\ \frac{1}{4a^2}\Tr H^2\right]\ , \ \ a\in[0,1]\ .
\label{P12}
\eea
We expect that the simplest, Gaussian choice that we have made here is
not important in the large-$n$ limit due to universality.
We can now define the following  $N_f$ flavour partition function
\be
{\cal Z}_{\nu}^{(N_f)}(m;\{z_f\};a)\sim
\int dH dW \prod_{f=1}^{N_f}
\det[\Dirac_5 +z_f\eins_N]P_1(W)P_2(H)\ ,
\label{Z2MM}
\ee
where we have inserted a product of $N_f$ characteristic polynomials
of real arguments $z_f$, in addition to the weight functions.
The integration is over all independent matrix elements of $W$ and $H$.

The role of the real parameter $0\leq a\leq1$ is to interpolate
between two different symmetry classes. For $N_f=0$ and $m=0$ the limit $a\to0$
leads to a delta function in all matrix elements of $H$, reducing
eq. (\ref{Z2MM}) to the chGUE. In the opposite limit $a\to1$ we obtain
a delta function in all matrix elements of $W$, and we are lead to the
GUE of size $N$. For $m\neq0$ we interpolate between the eigenvalues
of the Dirac operator with finite quark mass $m$ and the GUE coupled
to a fixed external field with eigenvalues $\pm m$, as it was
considered for example in \cite{Brezin}.

In \cite{DSV} a very similar Wilson chiral RMT was introduced which we will
label by $I\!I$. Starting from a non-Hermitian Wilson Dirac operator $D_W$,
\be
D_W\ =\
\left(\begin{array}{cc}
aA& W\\
-W^\dag&aB\\
\end{array}\right),
\ee
with $W$ as above and $A=A^\dag$ $n\times n$ and $B=B^\dag$
$(n+\nu)\times(n+\nu)$ Hermitian matrices, respectively,
the Hermitian Wilson Dirac operator $D_5$ in \cite{DSV}
was obtained by multiplying $D_W+m$ with $\gamma_5$:
\be
D_5=\left(\begin{array}{cc}
\eins_n& 0\\
0&-\eins_{n+\nu}\\
\end{array}\right)(D_W+m\eins_N)\ =\
\left(\begin{array}{cc}
m\eins_{n}& W\\
W^\dag&-m\eins_{n+\nu}\\
\end{array}\right)
+\left(\begin{array}{cc}
aA& 0\\
0&-aB\\
\end{array}\right),
\label{D5IIdef}
\ee
which is Hermitian. The corresponding partition function with Gaussian
weights reads
\be
{\cal Z}_{I\!I,\nu}^{(N_f)}(m;\{z_f\},a)\sim
\int dH dW \prod_{f=1}^{N_f}
\det[D_5
+z_f\eins_N]\exp\left[-\frac{1}{4}\Tr(A^2+B^2+2WW^\dag)\right]\ .
\label{Z3MM}
\ee
Because of symmetry the minus sign in the lower right corner in $B$ in
eq. (\ref{D5IIdef}) can be
absorbed by shifting $B\to-B$ in the integrand.

Both matrix models ${\cal Z}_{\nu}^{(N_f)}$ and ${\cal
  Z}_{I\!I,\nu}^{(N_f)}$  enjoy the same 
relationship to WchPT in the epsilon regime
in the following large-$N$ limit, stated here for equal arguments
$z_f=z$ (see appendix \ref{equiv} for the non-degenerate case): 
\be
\lim_{N\to\infty} {\cal Z}_{\nu}^{(N_f)}(m;z;a)=
\lim_{N\to\infty} {\cal Z}_{I\!I,\nu}^{(N_f)}(m;z;a)=
Z_\nu^{(N_f)}(\hat m,\hat z,\ha)\ ,
\label{match}
\ee
where
\be
Z_\nu^{(N_f)}(\hat m,\hat z,\hat a)=
\int_{U(N_f)} dU \det[U]^\nu \e^{\frac12 m_p \Sigma V\Tr(U+U^\dag)
+\frac12 z_p \Sigma V\Tr(U-U^\dag)
-{a_p}^2VW_8\Tr(U^2+U^{\dag\,2})}\ .
\label{eWchPT}
\ee
Here we have to match the large-$N$ with the large-$V$ infinite volume
limit\footnote{In \cite{DSV,ADSV} all matrix elements are rescaled by
  $\sqrt{N}$ to have compact support at $N=\infty$. This will modify
  these scaling relations on the RMT side by $\sqrt{N}$ for $z$ and $m$.}
\be
\hat m \equiv m_p\Sigma V= m\sqrt{N}\ ,\ \ \hat z\equiv z_p \Sigma V=
z\sqrt{N}\ ,
\ \ \mbox{and}\ \
\hat{a}^2\equiv a_p^2VW_8=\frac14 a^2 N\ .
\label{mza-rel}
\ee
While the equivalence eq. (\ref{match}) for the second RMT
eq. (\ref{Z3MM}) was shown in 
\cite{DSV} we will show this equivalence for our model
eq. (\ref{Z2MM}) in appendix \ref{equiv}.

Eq. (\ref{eWchPT}), the partition function of WchPT after Fourier
transform, usually contains
two further low-energy constants $W_6$ and $W_7$, apart from the
infinite volume chiral condensate $\Sigma$, and $W_8$ which encodes the
effects from a finite-lattice spacing to order ${\cal O}(a^2)$,
which we
displayed. Because $W_6$ and $W_7$ can be switched on at the expense
of one additional Gaussian integral each \cite{ADSV},
we will only consider $W_6=W_7=0$ in the following.
The parameter $m_p$ denotes the standard (equal) quark-mass term
coupling to $\bar \psi\psi$ in field theory,  whereas the
$z_p$ denotes an additional source terms coupling to $\bar \psi\gamma_5\psi$
which will be convenient later.

Because of the matching eq. (\ref{match}) we will assume that both
RMT eqs. (\ref{Z2MM}) and (\ref{Z3MM}) as well as WchPT
eq. (\ref{eWchPT}) are in the same universality class, also regarding
all eigenvalue density correlation functions.

The matching between the two matrix models can be made more explicit on the
level of matrix elements. By redefining $H/a\to H$ and
$W/\sqrt{1-a^2}\to W$ in eq. (\ref{D5Idef}) we obtain to leading order
\be
\Dirac_5 \ =
\left(\begin{array}{cc}
m\eins_n& W\\
W^\dag &-m\eins_{n+\nu}\\
\end{array}\right)+
\left(\begin{array}{cc}
aA & a\Om+\frac12 a^2W\\
a\Om^\dag-\frac12 a^2W^\dag &aB\\
\end{array}\right)+ {\cal O}(a^4)\ ,
\ee
where
\be
H=
\left(\begin{array}{cc}
A& \Om\\
\Om^\dag& B\\
\end{array}\right)\ .
\label{Hdef}
\ee
The advantage of allowing for extra,
off-diagonal matrix elements $\Om$ in matrix $H$ in our model is that
we can do a proper 
change of variables from $\{W,H\}$ to $\{W,\Dirac_5\}$, keeping the same
number of degrees of freedom. This allows us to go to an
eigenvalue basis as we will show in the next section.

Furthermore, the choice of
parametrisation in $a$ in our model is more convenient when
studying the symmetry transition from the chGUE at $a=0$ to the GUE at
$a=1$.
Compared to that in
$D_5$ eq. (\ref{D5IIdef}) the transition starting at $a=0$ corresponds to
the chGUE too, then leading via $a={\cal O}(1)$ of a
full $N\times N$ GUE to a decoupling for $a\gg1$ into two GUEs of
sizes $n$ and $n+\nu$ respectively.

\sect{The eigenvalue picture}\label{jpdf}

In this section we will derive an eigenvalue representation of our
partition function eq. (\ref{Z2MM}).
It can be obtained by the following change of variables from
$H$ to  $\Dirac_5$,
\be
H
=\
\Dirac_5-\left(\begin{array}{cc}
m\eins_n& W\\
W^\dag&-m\eins_{n+\nu}\\
\end{array}\right)\ ,
\label{change}
\ee
then diagonalising $\Dirac_5$ and $W$, using the HCIZ formula,
and then finally integrating out the $W$-eigenvalues.

The Jacobian of the change of variables eq. (\ref{change}) is a
constant and we get for the partition function
\bea
{\cal Z}_{\nu}^{(N_f)}(m;\{z_f\};a)&\sim& \int d\Dirac_5 dW
\prod_{f=1}^{N_f}\det[\Dirac_5 +z_f\eins_N]
P_1(W) P_2\left(\Dirac_5-\left(\begin{array}{cc}
m\eins_n& W\\
W^\dag&-m\eins_{n+\nu}\\
\end{array}\right)\right)
\nn\\
&=& \e^{-\ \frac{Nm^2}{4a^2}}
\int d\Dirac_5\prod_{f=1}^{N_f}\det[\Dirac_5 +z_f\eins_N]
\exp\Big[-\frac{1}{4a^2}\Tr[\Dirac_5^2]\Big]\nn\\
&&\times \int dW
\exp\left[
-\frac{1}{2a^2(1-a^2)}\Tr\Big[WW^\dag\Big]
+\frac{1}{2a^2}\Tr\left[
\Dirac_5\Big(\begin{array}{cc}
m\eins_n& W\\
W^\dag&-m\eins_{n+\nu}\\
\end{array}\Big)
\right]\right].
\nn\\
&&
\label{ZDW}
\eea
Now we can diagonalise $\Dirac_5$ by a unitary transformation
$U\in U(N)$,
\bea
\Dirac_5&=& UDU^\dag\ ,\ \ D=\mbox{diag}(d_1,\ldots,d_{n})\ ,\
d_j\in\mathbb{R}\ ,
\eea
and the chiral matrix $W$ by a
singular value decomposition (or equivalently a diagonalisation of $WW^\dag$):
\bea
W&=& uYv\
\Leftrightarrow
\left(\begin{array}{cc} m\eins_n& W\\W^\dag&-m\eins_{n+\nu}\\ \end{array}\right)
=
\left(\begin{array}{cc}
0& v^\dag\\
u&0\\
\end{array}\right)
\left(\begin{array}{cc}
m\eins_n& Y\\
Y^T&-m\eins_{n+\nu}\\
\end{array}\right)
\left(\begin{array}{cc}
0& u^\dag\\
v&0\\
\end{array}\right),
\eea
where $Y$ is a rectangular $n\times(n+\nu)$ matrix with real positive
diagonal elements $\{y_1,\ldots,y_{n}\}$, $y_j\geq0$.
Note the different numbers $N=2n+\nu$
and $n$ of these eigenvalues $d_j$ and $y_j$ , respectively.
Including the Jacobians of these
transformations, which contain the standard Vandermonde determinants,
\be
\Delta_{N}(\{d\})\equiv\prod_{j>l}^{N}(d_j-d_l)\ , \ \
\Delta_{n}(\{y^2\})=\prod_{j>l}^{n}(y_j^2-y_l^2),
\label{Deltadef}
\ee
we get
\bea
{\cal Z}_{\nu}^{(N_f)}(m;\{z_f\};a)&\sim&
\e^{-\ \frac{Nm^2}{4a^2}}
\int_{-\infty}^\infty\prod_{j=1}^{N}
dd_j\prod_{f=1}^{N_f}(d_j+z_f) \e^{-\frac{d_j^2}{4a^2}}
\Delta_{N}(\{d\})^2\ \int_0^\infty\prod_{b=1}^{n} dy_by_b^{2\nu+1}\
\e^{\frac{- y_b^2}{2a^2(1-a^2)}}
\nn\\
&&\times  \Delta_n(\{y^2\})^2\ 
\int dUdudv\ \exp\left[+\frac{1}{2a^2}\Tr\Big[
UDU^\dag \left(\begin{array}{cc}
m\eins_{n}& Y\\
Y^T&-m\eins_{n+\nu}\\
\end{array}\right)\Big]\right],
\label{ZpreHC}
\eea
where we have used the invariance of the Haar measure under the transformation
\be
U\to \left(\begin{array}{cc}
0& v^\dag\\
u&0\\
\end{array}\right)U\ .
\ee
The unitary integrations can now be carried out and we obtain from the
HCIZ integral
\be
\int dU \exp\left[\frac{1}{2a^2}
\Tr\Big[ UDU^\dag \left(\begin{array}{cc}
m\eins_{n}& Y\\
Y^T &-m\eins_{n+\nu}\\
\end{array}\right)\Big]\right]
\ =\ \frac{\det_{1\leq i,j\leq N}
\Big[\exp[\frac{1}{2a^2}d_iw_j]\Big]}{\Delta_{N}(\{d\})
\Delta_{N}(\{w\})}
\ .
\label{HCIZ}
\ee
Here $w_k$, $k=1,\ldots,N$ are the $N$ eigenvalues of the matrix
$\left(\begin{array}{cc}
m\eins_{n}& Y\\
Y^T&-m\eins_{n+\nu}\\
\end{array}\right)
$
which
we will now relate to the eigenvalues $y_b$ in order to partly cancel
the Vandermonde determinants.
From the solutions of the
characteristic equation
\bea
0&=&\det\left[\la\eins_{N}-\left(\begin{array}{cc}
m\eins_{n}& Y\\
Y^T&-m\eins_{n+\nu}\\
\end{array}\right)\right]
= (\la+m)^\nu\prod_{b=1}^{n}\Big( \la^2-m^2-y_b^2\Big)
\ =\ \prod_{j=1}^{N}(\la-w_j)\ ,
\eea
the eigenvalues $w_k$ are given by\footnote{From this it can be seen
  that at $a=0$ there is an accumulation of eigenvalues at $-m$. Note
  that our convention differs from \cite{DSV,ADSV}.}
\bea
w_j&=&\ \
\sqrt{m^2+y_j^2\ }\ ,\ j=1,\ldots,n\nn\\
w_{j+n}&=&-\
\sqrt{m^2+y_{j}^2}\ ,\
j=1,\ldots,n\nn\\
w_{j+2n}&=& -m+\eps_{j}\ ,\ j=1,\ldots,\nu\ .
\label{wdef}
\eea
Here we have lifted the degeneracy of the last $\nu$
eigenvalues by adding small
parameters $\eps_1,\ldots,\eps_{\nu}$ which will be sent to zero
later (for $\nu=1$ this is not necessary, but we will keep $\eps_1\neq0$ for
symmetry reasons). Making the $\eps$-dependence explicit in the Vandermonde
we can write
\bea
\Delta_{N}(w)&=&\Delta_\nu(\{\eps-m\}) \prod_{l=1}^{2n}\prod_{h=1}^\nu
(\eps_h-m-w_l)\
 \prod_{i>j}^{n}(w_{i+n}-w_{j+n})(w_i-w_j)
\prod_{o,q=1}^{n}(w_{o+n}-w_q)\nn\\
&=&(-)^{\frac12 n(3n-1)}2^n\Delta_\nu(\{\eps\}) \prod_{l=1}^{n}\prod_{h=1}^\nu
(\eps_h^2-2m\eps_h-y^2_l)\  \Delta_{n}(\{y^2\})^2
\  \prod_{i=1}^{n}\sqrt{m^2+y_i^2}\ ,
\label{Deltasplit}
\eea
after some algebra.
The first Vandermonde determinant $\Delta_\nu(\{\eps\})$ will cancel
the zeros from the numerator in eq. (\ref{HCIZ})
when we take the limit of degenerate eigenvalues $\eps_k\to0$, whereas in
the remaining factors in eq. (\ref{Deltasplit}) this limit is smooth.

Using induction one can easily obtain the following limiting result.
For $\nu=3$ we have to Taylor expand
the next column to
one order higher, when taking $\eps_3\to\eps_1$, etc.,
and so we end up with the following result to leading order
\bea
&&\lim_{\eps_k\to0}\frac{1}{\Delta_\nu(\{\eps\})}
\left|
\begin{array}{cccccc}
\e^{x_1w_1}     &\ldots&\e^{x_1w_{2n}}
                &\e^{x_1(-m+\eps_1)}&\cdots&\e^{x_1(-m+\eps_\nu)}\\
\cdots          &\cdots&             \cdots& \cdots&\cdots&\cdots\\
\e^{x_Nw_1}&\ldots&\e^{x_Nw_{2n}}&\e^{x_N(-m+\eps_\nu)}&\cdots&
\e^{x_N(-m+\eps_\nu)}\\
\end{array}
\right|\\
&&=\prod_{j=0}^{\nu-1}\frac{1}{j!}
\left|
\begin{array}{ccccccc}
\e^{x_1w_1}     &\ldots&\e^{x_1w_{2n}}&\e^{-x_1m}&x_1\e^{-x_1m}&\cdots
&x_1^{\nu-1}\e^{-x_1m}\\
\cdots          &\cdots&  \cdots          &\cdots &\cdots&\cdots& \cdots\\
\e^{x_{N}w_1}&\ldots&\e^{x_Nw_{2n}}&\e^{-x_{N}m}&x_{N}\e^{-x_{N}m}
&\cdots&x_{N}^{\nu-1}\e^{-x_{N}m}\\
\end{array}\right|,\nn
\eea
where we have defined $x_{i}=\frac{1}{2a^2}d_i$ for $i=1,\ldots,N$
for later use.

We can now write the full answer of our partition function eq. (\ref{ZpreHC})
in term of the two sets of eigenvalues only:
\bea
{\cal Z}_{\nu}^{(N_f)}(m;\{z_f\};a)\!&\sim&
\!\e^{-\frac{-m^2(2n+N(1-a^2))}{4a^2(1-a^2)}}
\int_{-\infty}^\infty\prod_{j=1}^{N}
dd_j\prod_{f=1}^{N_f}(d_j+z_f) \e^{-\frac{d_j^2}{4a^2}}
\Delta_{N}(\{d\})
\int_m^\infty\prod_{b=1}^{n} du_b
\e^{\frac{-u_b^2}{2a^2(1-a^2)}}
\nn\\
&&\times
\left|
\begin{array}{cccc}
\e^{x_1u_1}\ldots\e^{x_1u_n}&\e^{-x_1u_1}\ldots\e^{-x_1u_n}&\e^{-x_1m}&x_1\,
\e^{-x_1m}\ldots 
x_1^{\nu-1}\ \e^{-x_1m}\\
\cdots &\cdots&\cdots& \cdots\\
\e^{x_{N}u_1}\ldots\e^{x_Nu_n}&\e^{-x_{N}u_1}\ldots\e^{-x_Nu_n}&
\e^{-x_{N}m}&x_{N}\e^{-x_{N}m} 
\ldots x_{N}^{\nu-1}\e^{-x_{N}m}\\
\end{array}\right|,\nn\\
\label{Zev}
\eea
after dropping $m$-independent constants and changing variables
$u_j=\sqrt{m^2+y_j^2}$. For $\nu=0$ the last $\nu$ columns that are
independent of $u_j$ have to be dropped.

We will now apply a generalisation of the de Bruijn integration formula
in order to integrate out the set of variables $u_k$.
It is given by \cite{KG} (see appendix C.2 therein)
\bea
&&\prod_{q=1}^{n} \int du_q w(u_q)
\det_{1\leq b\leq 2n+\nu;\,1\leq j\leq n;\,1\leq i\leq \nu}
\Big[\{\phi_b(u_{j}),\psi_b(u_{j})\}\ A_{bi}\ \Big]\nn\\
=&&
(-)^{\nu(\nu-1)/2}n!\underset{1\leq b,c\leq
  2n+\nu;\,1\leq i\leq \nu}{\Pf}\left[\!\!
\begin{array}{ll}
\left\{\int du \,
w(u)[\phi_b(u)\psi_c(u)-\phi_c(u)\psi_b(u)]
\right\}\!&\!A_{bi}\\
-A^T_{ic}&\!0\\
\end{array}\!
\right].
\nn\\
&&
\label{gendeB}
\eea
Here $A_{bi}=d_b^{i-1}\e^{-d_bm/2a^2}$ is a matrix of size
$N\times\nu$. When $\nu=0$ it is 
absent and we are back to the standard de Bruijn formula.
In our case we have
\bea
{\cal Z}_{\nu}^{(N_f)}(m;\{z_f\};a) &\sim&
\exp\Big[\frac{Nm^{2}}{2a^2(1-a^2)}\Big]
\int_{-\infty}^\infty\prod_{j=1}^{2n+\nu}
dd_j\prod_{f=1}^{N_f}(d_j+z_f)
\exp\Big[{-\frac{d_j^2}{4a^2}}\Big]
\Delta_{2n+\nu}(\{d\})\ \ \ \ 
\label{ZDevPf}
\\
&&\times\Pf_{1\leq i,j\leq 2n+\nu;\,1\leq q\leq \nu}
\left[
\begin{array}{ll}
F(d_j-d_i)& \!d_i^{\, q-1}\e^{-\frac{d_im}{2a^2}}\\
 -d_j^{\, q-1}\e^{-\frac{d_jm}{2a^2}} & {\mathbf 0}_{\nu\times\nu}\\
\end{array}
\right]\ ,
\nn
\eea
where we have defined the antisymmetric weight
\bea
F(x)
&\equiv& \e^{\frac{x^2(1-a^2)}{8a^2}}
\left[\erf\Big(\frac{x\sqrt{1-a^2}}{\sqrt{8a^2}}
+\frac{m}{\sqrt{2a^2(1-a^2)}}\Big)  
+
\erf\Big(\frac{x\sqrt{1-a^2}}{\sqrt{8a^2}}-\frac{m}{\sqrt{2a^2(1-a^2)}}\Big)
\right]
\label{Fdef}\\
&=& \frac{4}{\sqrt{2\pi a^2(1-a^2)}}\int_m^\infty du\
\e^{-\frac{u^2}{2a^2(1-a^2)}} 
\sinh\Big[\frac{xu}{2a^2}\Big].\nn
\eea
We have also used the antisymmetry of 
$\erf(x)=\frac{2}{\sqrt{\pi}}\int_0^xds\,e^{-s^2}=-\erf(-x)$.
Eq. (\ref{ZDevPf}) is the partition function in terms of the joint
probability distribution functions (jpdf) of its eigenvalues.

Let us give the two examples that we will solve explicitly in the
following sections. For $N_f=\nu=0$ we have
\bea
{\cal Z}_{\nu=0}^{(N_f=0)}(m;a)
\equiv
\int_{-\infty}^\infty\prod_{j=1}^{2n}
dd_j \exp\Big[-\frac{d_j^2}{4a^2}\Big]
\ \Delta_{2n}(\{d\})\Pf_{1\leq i,j\leq 2n}[F(d_j-d_i)]\ .
\label{Zjpdf}
\eea
Here we have omitted all constants that will drop out in expectation
values and density correlation functions to be considered in the next section.
For $\nu=1$ and $N_f=0$ we have
\bea
\label{ZDevPf+1}
{\cal Z}_{\nu=1}^{(N_f=0)}(m;a)
\equiv
\int_{-\infty}^\infty\prod_{j=1}^{2n+1}
dd_j
\exp\Big[{-\frac{d_j^2}{4a^2}}\Big]
\Delta_{2n+1}(\{d\})
\Pf_{1\leq i,j\leq 2n+1}
\left[
\begin{array}{ll}
F(d_j-d_i)& \e^{-\frac{d_i m}{2a^2}}\\
-\e^{-\frac{d_j m}{2a^2}}  & 0\\
\end{array}
\right]\!.
\eea

Let us add one remark on the universality of our model. In order to be
able to use the HCIZ formula it was crucial that we started with a
Gaussian distribution of matrix elements. However, after having arrived
at an eigenvalue representation eq. (\ref{ZDevPf}) we could take this
as a starting point and we would obtain the same results below,
if we were to replace the Gaussian distribution $\exp[-d_j^2/4a^2]$ by a
more general weight function $w(d_j)$. The same general framework to compute
density correlation functions to be presented in the next section
could be applied.

\sect{All density correlation functions for finite-$N$}\label{Rk}

The partition function eq. (\ref{Zjpdf}) is very much reminiscent of the one
for the GUE-GOE transition in \cite{PandeyMehta}, apart from the
different function $F(x)$ inside the Pfaffian. We can therefore apply
the general method of solving this two-matrix model developed in
\cite{NF98} for general weight functions, which applies equally to
eq. (\ref{ZDevPf+1}) with $\nu=1$. From now on we will restrict
ourselves to $\nu=0,1$ and $N_f=0$.

The $k$-point density correlation functions that we will determine are
defined as follows for $\nu=0$
\be
\rho_k(d_1,\ldots,d_k)\equiv\frac{1}{{\cal Z}_{\nu=0}^{(0)}}\frac{N!}{(N-k)!}
\int_{-\infty}^\infty dd_{k+1}\ldots dd_N
\prod_{j=1}^{N} w(d_j)
\ \Delta_{N}(\{d\})\Pf_{1\leq i,j\leq N}[F(d_j-d_i)]\ ,
\label{rhokdef}
\ee
with even $N=2n$, and for $\nu=1$ as 
\be
\rho_k^{\nu=1}(d_1,\cdots,d_k) = \frac{1}{{\cal Z}^{(0)}_{\nu = 1}} 
\frac{N!}{(N-k)!} 
\int_{-\infty}^{\infty} dd_{k+1} \cdots dd_N  
\prod_{j=1}^N w(d_j) \Delta_N(\{d\}) {\rm Pf}_{1 \leq i,j \leq
  N}\!\!\left[ \!\!\!
\begin{array}{cc}  
F(d_j - d_i) & f(d_i) \\ - f(d_j) & 0 \end{array} 
\!\!\!
\right] ,
\ee
with odd $N = 2 n + 1$. 
Here we have used the abbreviations 
\be
w(x)\equiv\exp[-x^2/(4a^2)]\ ,\ \ f(x) \equiv \exp[- m x/(2 a^2)]\ .
\label{weight}
\ee
The expressions for the solution in terms of three kernels will 
hold for a general weight functions $w(x),\ f(x)$  and $F(x)$ though.
Here and in the following we will drop the label $\nu=0$ and only
index $\nu=1$. Furthermore we also suppress the
dependence on the parameters $m,a$ in the
arguments for simplicity.

The correlation functions defined above 
can be expressed as the Pfaffian of the
following matrix \cite{NF98}
\be
\rho^\nu_k(d_1,\ldots,d_k)= \Pf_{1\leq i,j\leq k}
\left[
\begin{array}{cc}
I^\nu_n(d_i,d_j)& S^\nu_n(d_i,d_j)\\
-S^\nu_n(d_j,d_i)& -D^\nu_n(d_i,d_j)\\
\end{array}
\right],
\label{rhoPf}
\ee
given in terms of three kernels. 

Let us begin with the case $\nu=0$ where we have \cite{NF98}
\bea
S_n(x,y)&=& \sum_{j=1}^n\frac{w(y)}{r_{j-1}}
\Big(\phi_{2j-2}(x)R_{2j-1}(y)-\phi_{2j-1}(x)R_{2j-2}(y)
\Big)\ ,\nn\\
D_n(x,y)&=& \sum_{j=1}^n\frac{w(x)w(y)}{r_{j-1}}
\Big(R_{2j-2}(x)R_{2j-1}(y)-R_{2j-1}(x)R_{2j-2}(y)
\Big)\ ,\nn\\
I_n(x,y)&=&- \sum_{j=1}^n\frac{1}{r_{j-1}}
\Big(\phi_{2j-2}(x)\phi_{2j-1}(y)-\phi_{2j-1}(x)\phi_{2j-2}(y)
\Big)\ -\ F(x-y)\ .
\label{3kernels}
\eea
In our specific case of weights eqs. (\ref{Fdef}) and (\ref{weight})
these kernels are given by the following monic
skew-orthogonal polynomials $R_j(x)=x^j+\ldots$ and their integral transforms,
\bea
R_{2j}(x)&=&\frac{j!\ 2^j(1-a^2)^j}{(-1)^j\sqrt{\pi}}
\int_{-\infty}^\infty ds\ \e^{-s^2}L_j\left(
  \frac{(x+2ias)^2-m^2}{2(1-a^2)}\right) , \label{sOPs}\\
R_{2j+1}(x)&=&-2a^2w(x)^{-1}
\frac{d}{dx}\Big[
w(x)R_{2j}(x)\Big]\nn\\
&=&\frac{j!\ 2^j(1-a^2)^j}{(-1)^j\sqrt{\pi}}
\int_{-\infty}^\infty ds\ \e^{-s^2}(x+2ias)L_j\left(
  \frac{(x+2ias)^2-m^2}{2(1-a^2)}\right),
\nn\\
\phi_j(x)&=&\int_{-\infty}^\infty dy\ w(y)F(x-y)R_j(y)\ .
\label{trafo}
\eea
Here we have used an integration by parts to simplify the expression
for the odd polynomials. The polynomials have 
parity, $R_j(-x)=(-)^jR_j(x)$, as can be easily seen from the integral
representations.
These skew-orthogonal polynomials satisfy the following relations as
it is shown in appendix \ref{SOP-C} in detail:
\bea
\langle R_{2j}, R_{2l+1}\rangle=r_j\delta_{jl}\ ,\nn\\
\langle R_{2j}, R_{2l}\rangle=0=\langle R_{2j+1}, R_{2l+1}\rangle
\label{skewdef}
\eea
with the skew product defined as
\be
\langle h,g\rangle\equiv \int_{-\infty}^\infty dxdy\,
w(x)w(y)F(x-y)h(y)g(x)\ .
\label{skew}
\ee
The squared norms read
\be
r_j\equiv\langle R_{2j}, R_{2j+1}\rangle=8\sqrt{2\pi a^2}
(1-a^2)^{2j+\frac12}2^{2j}
(j!)^2 \e^{-\ \frac{m^2}{2(1-a^2)}}\ ,
\label{norm}
\ee
are also derived in 
appendix \ref{SOP-C}. The quenched partition function can now
be expressed in term of these norms and is given by
\be
{\cal Z}^{(N_f=0)}_{\nu=0}=N! \prod_{j=0}^{n-1}r_j\ .
\ee

Because of the special form of our polynomials the kernels can be
simplified. We start with the kernel $D_n(x,y)$, the other two easily
follow from the relations 
\bea
S_n(x,y)&=& \int_{-\infty}^\infty dz\ F(x-z) D_n(z,y)\ ,\nn\\
I_n(x,y)+F(x-y)&=& - \int_{-\infty}^\infty dz\ F(x-z) S_n(x,z)\ .
\label{kernelrel}
\eea
Inserting eq. (\ref{sOPs}) into the definition we obtain
\bea
D_n(x,y)&=&\e^{-\frac{x^2+y^2}{4a^2}}
\frac{\e^{\frac{m^2}{2(1-a^2)}}}{8\pi\sqrt{2\pi    a^2(1-a^2)}} 
\int_{-\infty}^\infty ds\, dr\ \e^{-s^2-r^2}[(y+2ias)-(x+2iar)]\nn\\
&&\times\sum_{j=0}^{n-1}L_j\left( \frac{(x+2iar)^2-m^2}{2(1-a^2)}\right)
L_j\left(\frac{(y+2ias)^2-m^2}{2(1-a^2)}\right)\ .
\label{D1}
\eea
Using the Christoffel-Darboux identity
\be
\sum_{j=0}^{n-1}L_j(X)L_j(Y)=-n\
\frac{L_n(X)L_{n-1}(Y)-L_n(Y)L_{n-1}(X)}{X-Y}\ ,
\ee
we obtain the following final expression
\bea
D_n(x,y)&=&\e^{-\frac{x^2+y^2}{4a^2}}
\frac{\e^{\frac{m^2}{2(1-a^2)}}n\sqrt{1-a^2}}{4\pi\sqrt{2\pi a^2}}
\int_{-\infty}^\infty ds\, dr\ \e^{-s^2-r^2}
\label{Dnfinal}\\
&&\times
\frac{L_n\left(\frac{(x+2iar)^2-m^2}{2(1-a^2)}\right)
L_{n-1}\left(\frac{(y+2ias)^2-m^2}{2(1-a^2)}\right)
- L_{n-1}\left(\frac{(x+2iar)^2-m^2}{2(1-a^2)}\right)
L_{n}\left(\frac{(y+2ias)^2-m^2}{2(1-a^2)}\right)
}{(y+2ias)+(x+2iar)} .\nn
\eea
The simplest examples for the correlation functions at finite-$N$ are
the spectral 
density and the spectral two-point function. These are given by
\bea
\rho_1(x)&=&S_n(x,x)=\int_{-\infty}^\infty dz\ F(x-z) D_n(z,x)\ ,
\nn\\
\rho_2(x,y)&=&S_n(x,x)S_n(y,y)+I_n(x,y)D_n(x,y)-S_n(x,y)S_n(y,x)\ ,
\label{rho1}
\eea
with the insertion of eq. (\ref{Dnfinal}) and (\ref{kernelrel}), see
eq. (\ref{prelim}) below. This
ends the general solution of our model for finite-$N$ for $\nu=0$.\\

For $\nu=1$ the skew-orthogonal polynomials can be expressed in terms
of those for $\nu=0$. Following \cite{NF98} we introduce new polynomials
\begin{equation}
R^{\nu=1}_j(x) \equiv R_j(x) - \frac{s_j}{s_{N-1}} R_{N-1}(x), \ \ \  
j = 0,1,\cdots,N-2\ ,
\label{Rrel}
\end{equation}
and for the largest polynomial
\begin{equation}
R^{\nu=1}_{N-1}(x) \equiv R_{N-1}(x)\ ,
\end{equation}
with coefficients
\begin{equation}
s_j = \int_{-\infty}^{\infty} dx \ w(x) f(x) R_j(x).
\end{equation}
Moreover we define the integral transforms of the polynomials as in the
previous case eq. (\ref{trafo})
\begin{equation}
\phi^{\nu=1}_j(x) \equiv \int_{-\infty}^{\infty} dy \ F(x-y) w(y) 
R^{\nu=1}_j(y).
\end{equation}
The kernels for $\nu=1$ to be inserted in eq. (\ref{rhoPf}) read \cite{NF98}
\begin{eqnarray}
S^{\nu=1}_n(x,y) & = & \sum_{j=1}^n \frac{w(y)}{r_{j-1}} \left( 
\phi^{\nu=1}_{2 j - 2}(x) R^{\nu=1}_{2 j - 1}(y) - 
\phi^{\nu=1}_{2 j - 1}(x) R^{\nu=1}_{2 j - 2}(y) \right)
\nonumber \\ & & 
+ \frac{1}{s_{N-1}} f(x) w(y) R^{\nu=1}_{N-1}(y), \nonumber \\ 
D^{\nu=1}_n(x,y) & = & \sum_{j=1}^n \frac{w(x) w(y)}{r_{j-1}} \left( 
R^{\nu=1}_{2 j - 2}(x) R^{\nu=1}_{2 j - 1}(y) - 
R^{\nu=1}_{2 j - 1}(x) R^{\nu=1}_{2 j - 2}(y) \right),
\nonumber \\ 
I^{\nu=1}_n(x,y) & = & - \sum_{j=1}^n \frac{1}{r_{j-1}} \left( 
\phi^{\nu=1}_{2 j - 2}(x) \phi^{\nu=1}_{2 j - 1}(y) - 
\phi^{\nu=1}_{2 j - 1}(x) \phi^{\nu=1}_{2 j - 2}(y) \right) 
\nonumber \\ & & + \frac{1}{s_{N-1}} \left( \phi^{\nu=1}_{N-1}(x) f(y) - 
\phi^{\nu=1}_{N-1}(y) f(x) \right) - F(x - y).
\label{nu=1Ker}
\end{eqnarray}
The partition function is also evaluated as
\begin{equation}
{\cal Z}^{(N_f = 0)}_{\nu = 1} = N!\ s_{N-1} \prod_{j=0}^{n-1} r_j.
\end{equation}
As a last step we still have to determine the new coefficients $s_j$.
We can readily find
\begin{eqnarray}
s_{2 j} & = & \int_{-\infty}^{\infty} dx \ {\rm e}^{-x(x + 2 m)/(4 a^2)} 
R_{2 j}(x) \nonumber \\ 
& = & \frac{j! 2^j (1 - a^2)^j {\rm e}^{m^2/(4 a^2)}}{(-1)^j \sqrt{\pi}} 
\int_{-\infty}^{\infty} dx ds \ {\rm e}^{-(x + m)^2/(4 a^2) - s^2} L_j\left( 
\frac{(x + 2 i a s)^2 - m^2}{2 (1 - a^2)} \right) 
\nonumber \\ 
& = & (-1)^j \ j! \ 2^{j+1} (1 - a^2)^j \sqrt{\pi} 
\ a \ {\rm e}^{m^2/(4 a^2)}\ ,
\end{eqnarray}
for the even coefficients, using an argument as in eq. (\ref{normint}), and
\begin{eqnarray}
s_{2 j + 1} & = & - 2 a^2 \int_{-\infty}^{\infty} dx {\rm e}^{- m x/(2 a^2)} 
\frac{d}{dx}\left[ {\rm e}^{-x^2/(4 a^2)} R_{2 j}(x) \right] \nonumber \\ 
& = & - m \int_{-\infty}^{\infty} dx \ 
{\rm e}^{-x(x + 2 m)/(4 a^2)} R_{2 j}(x) \nonumber \\ 
& = & - m \ s_{2 j}\ ,
\end{eqnarray}
for the odd coefficients. This ends our general solution for $\nu=1$.

In order to check that the limit $a\to0$ correctly reproduces the
chGUE let us spell out the densities explicitly, where we start with
$\nu=0$. Here we use 
eq. (\ref{D1}) rather than eq. (\ref{Dnfinal}) and we obtain
\bea
\rho_1(x)&=&\int_{-\infty}^\infty dz
\frac{\e^{-\frac{(z+x)^2}{8a^2}+\frac{(z-x)^2}{8}+\frac{m^2}{2(1-a^2)}}}{8\pi
  a\sqrt{2\pi(1-a^2)}} 
\left[\erf\Big(
\frac{(x-z)(1-a^2)+2m}{2a\sqrt{2(1-a^2)}}
\Big)
+
\erf\Big(
\frac{(x-z)(1-a^2)-2m}{2a\sqrt{2(1-a^2)}}
\Big)
\right]\nn\\
&\times&\int_{-\infty}^\infty dsdr\, \e^{-s^2-r^2}(x-z+2ia(s-r))
\sum_{j=0}^{n-1}L_j\left( \frac{(z+2iar)^2-m^2}{2(1-a^2)}\right)
L_j\left(\frac{(x+2ias)^2-m^2}{2(1-a^2)}\right).\nn\\
&&
\label{prelim}
\eea 
Because of the delta-function that we obtain from 
\be
\lim_{a\to0}\frac{1}{2a\sqrt{2\pi}}\e^{-\frac{(z+x)^2}{8a^2}}=\delta(z+x)\ ,
\ee
we get the following from the error-functions (cf. \ref{limerf})
\be
\lim_{a\to0}\erf\left(
\frac{2x(1-a^2)\pm2m}{2a\sqrt{2(1-a^2)}}\right)=\sign(x\pm m)\ .
\ee
The resulting expression we can rewrite as 
\be
x(\sign(x+m)+\sign(x-m))=|x|2\Theta(|x|-m)\ .
\ee
In the second line in eq. (\ref{prelim}) the integrals decouple and we
obtain a simple sum of Laguerre polynomials. The final answer is thus 
\bea
\lim_{a\to0}\rho_1(x)&=& |x|\,\e^{-\frac{x^2}{2}+\frac{m^2}{2}}
\sum_{j=0}^{n-1}L_j\left(\frac{x^2-m^2}{2}\right)^2\ 
\Theta(|x|-m)
\label{rhonu0lim}\\
&=&\frac{|x|}{\sqrt{x^2-m^2}}\
\rho_1^{chGU\!E}\left(\sqrt{x^2-m^2}\right)\ 
\Theta(|x|-m)\ ,
\label{shift}
\eea
which corresponds to the shifted density of the chGUE for finite-$n$, due to the
extra mass in our Dirac operator definition eq. (\ref{D5Idef}),
\be
\rho_1^{chGU\!E}(y)= |y|\ \e^{-y^2/2}\sum_{l=0}^{n-1}L_l(y^2/2)^2\ .
\label{rhochGUE}
\ee

\begin{figure}[-h]
\centerline{\epsfig{figure=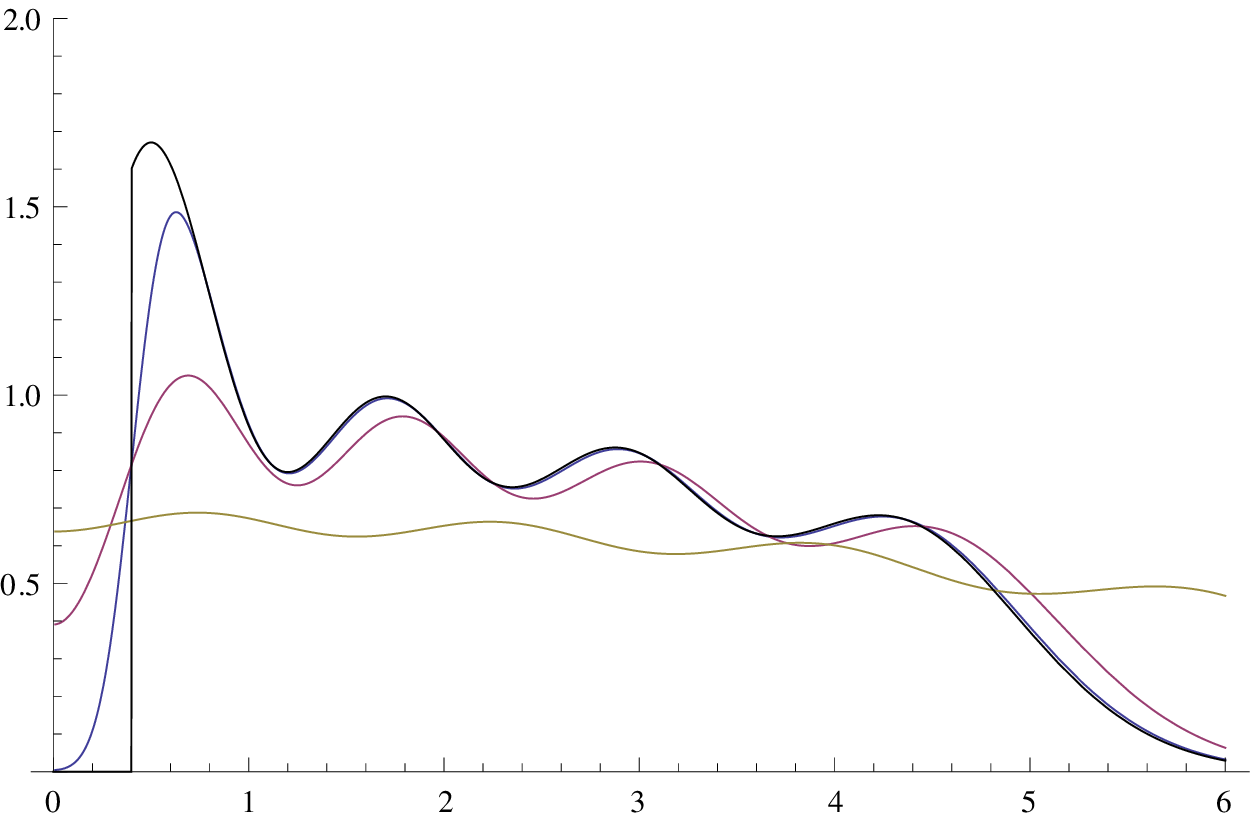,width=15pc}
\epsfig{figure=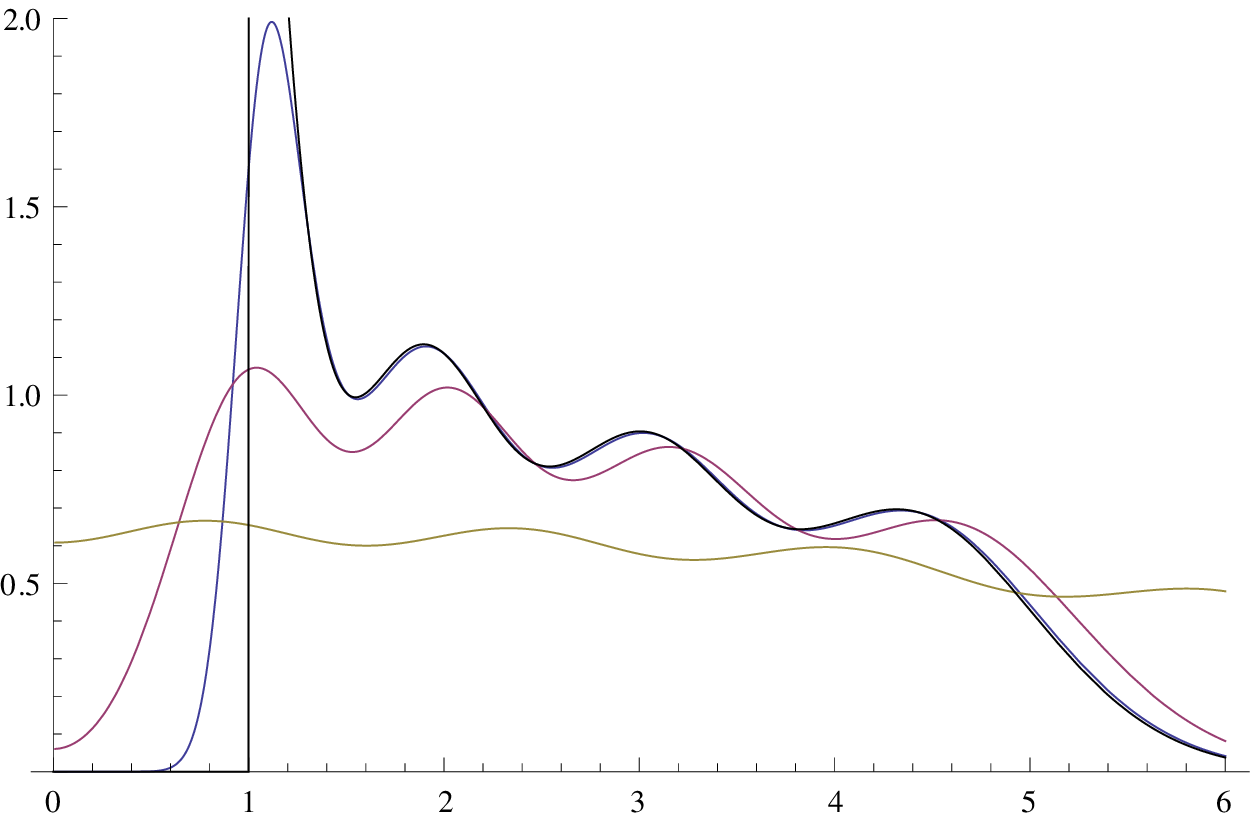,width=15pc}
}
\centerline{\epsfig{figure=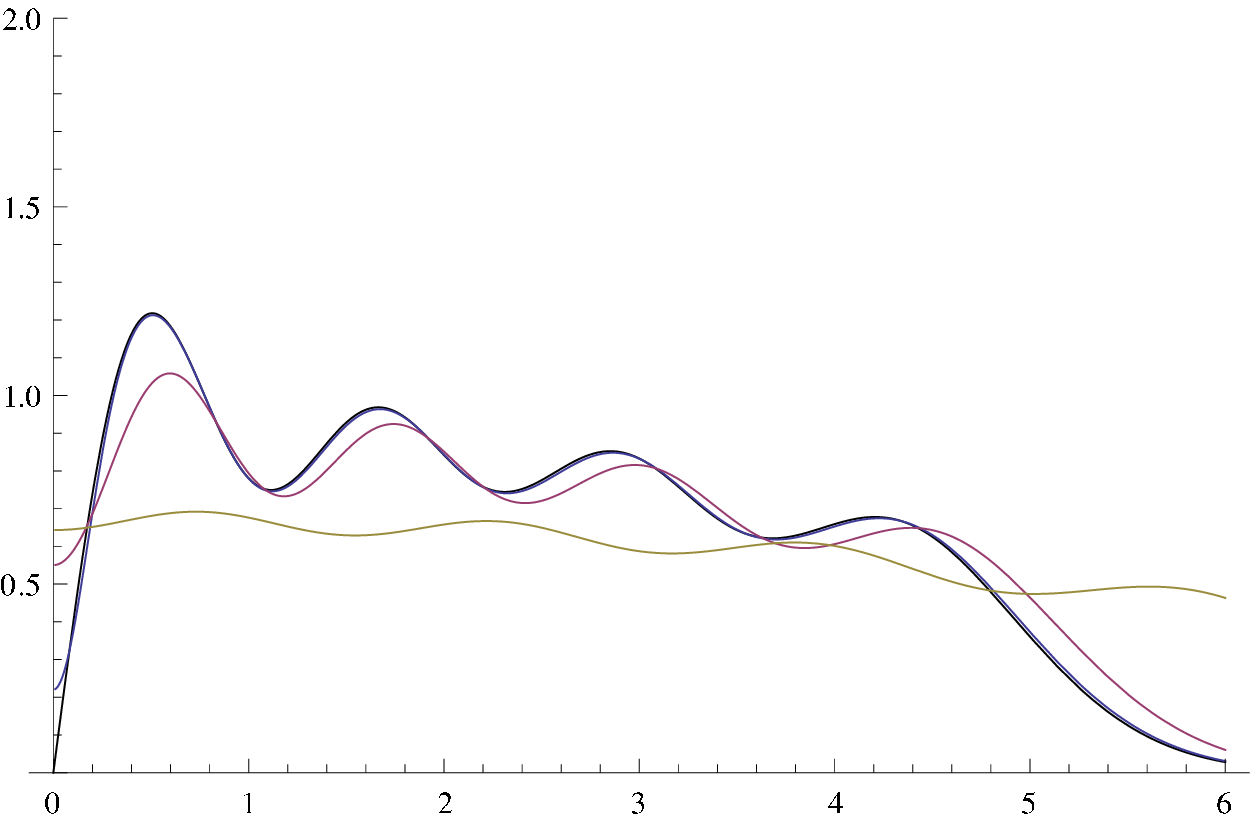,width=15pc}
}
\caption{Example for the $\nu=0$ quenched
density at finite $n=4$ at different masses: $m=0.4$ (top left plots)
and $m=1$ (top right plots):
the top black curves correspond to the shifted chGUE density
eq. (\ref{rhonu0lim}) at $a=0$ (which is clipped for $m=1$) ending at
a sharp cutoff for $m>0$, see eq. (\ref{shift}). The density
$\rho_1(x)$ eq. (\ref{prelim}) is plotted  
at various values of the deformation parameter $a$: 
at $a=0.1$ (top smooth blue curve),
$a=0.3$ (middle smooth red curve), and 
at $a=0.9$ (lowest yellow curve) which is already indistinguishable
from the GUE curve. Because the
spectral density is symmetric due to $\nu=0$ we only show the positive
eigenvalues. 
The lower plots show the same curves at $m=0$, the chGUE-GUE
transition. Here $a=0$ and
$a=0.1$ are already indistinguishable.
}
\label{rho1n4}
\end{figure}

As an illustration we have plotted the densities in figure \ref{rho1n4}.
For comparison we also display the finite-$n$ density of the chGUE
eq. (\ref{rhochGUE}) in the limit $a\to0$.
Switching on $a>0$ the smoothening of the hard wall provided by the
Heaviside-Theta functions is nicely seen in figure \ref{rho1n4}. 

\begin{figure}[-h]
\centerline{\epsfig{figure=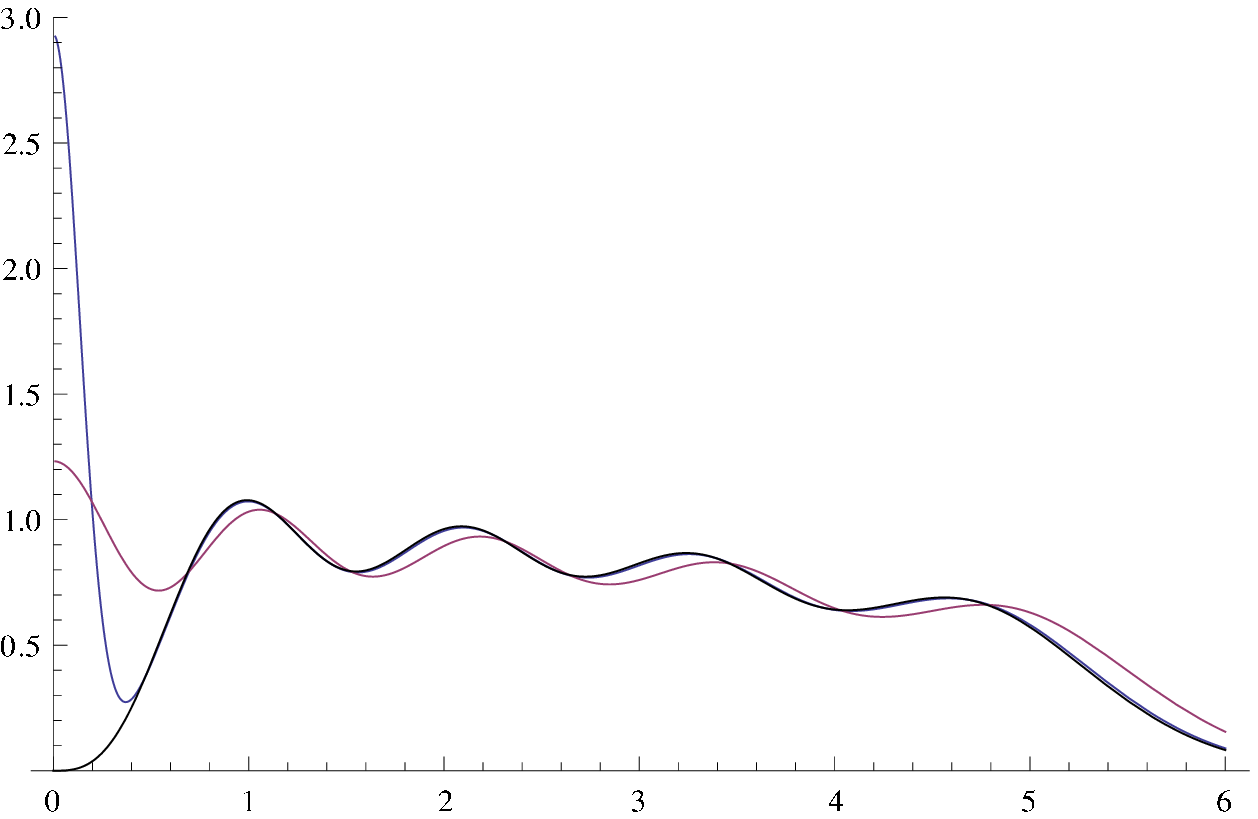,width=15pc}
\epsfig{figure=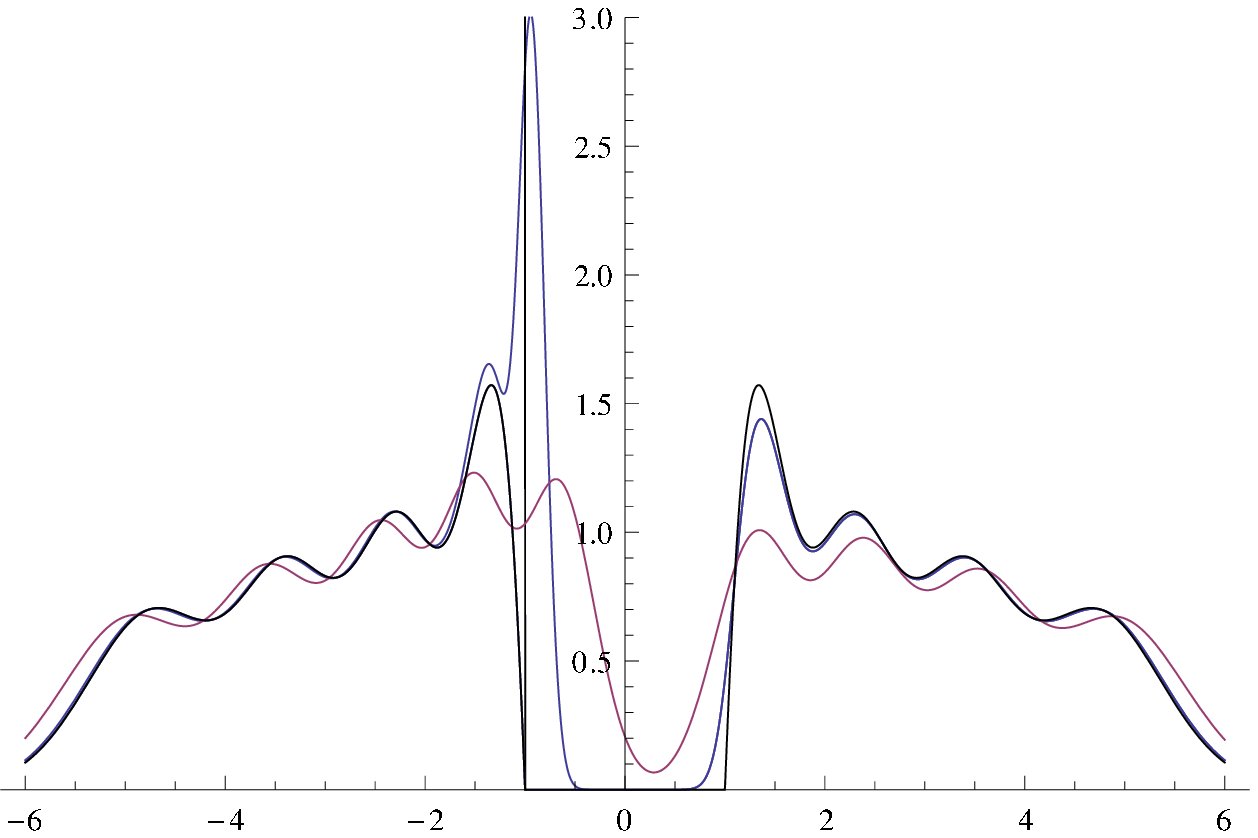,width=15pc}
}
\caption{Example for the $\nu=1$ quenched
density at finite $n=4$ at different masses: $m=0$ (left plots)
and $m=1$ (right plots). Because for $m=0$ the density is still
symmetric we only plot the positive eigenvalues. 
The black curves correspond to the (shifted) chGUE density
at $a=0$. It can be nicely seen how the delta-function at $x=-m$ gets
gradually smoothed out, where we show $a=0.1$ (blue curve) and
$a=0.3$ (red curve).
}
\label{rho1n4nu1}
\end{figure}

Let us now spell out the density for $\nu=1$ explicitly and perform
the limit $a\to0$ there. In view of the relation between the
polynomials eq. (\ref{Rrel}) it is useful to express the kernels for
$\nu=1$ in terms of those with $\nu=0$ as well. We begin with the
simplest kernel containing only polynomials and no integral transforms. After
some algebra we obtain
\bea
D^{\nu=1}_n(x,y) & = & D_n(x,y)+\e^{-\frac{(x^2+y^2)}{4a^2}}
\e^{\frac{m^2}{2(1-a^2)}}\frac{1}{8a\pi\sqrt{2\pi(1-a^2)}} 
\left\{
\int_{-\infty}^\infty ds\ \e^{-s^2}
L_n\left(\frac{(y+2ias)^2-m^2}{2(1-a^2)}\right)
\right.
\nn\\
&&\times
\left.
\int_{-\infty}^\infty dr\ \e^{-r^2}(x+2iar+m)
L^{(1)}_{n-1}\left(\frac{(x+2iar)^2-m^2}{2(1-a^2)}\right)
\ -\ (x\leftrightarrow y)
\right\}\ ,
\label{Dnu1}
\eea
where we used the following identity for generalised Laguerre
polynomials \cite{Grad} 
\be
\sum_{j=0}^{n-1}L_j(z)=L^{(1)}_{n-1}(z)\ .
\ee
For the kernel determining the density we obtain after inserting the
definitions
\bea
S^{\nu=1}_n(x,y) & = & \int_{-\infty}^\infty dz F(x-z)D^{\nu=1}_n(z,y)
+ \e^{-\frac{y^2+2xm+m^2}{4a^2}} \frac{1}{2\pi a}
\int_{-\infty}^\infty ds\
\e^{-s^2}L_n\left(\frac{(y+2ias)^2-m^2}{2(1-a^2)}\right).
\nn\\
\label{Snu1}
\eea
These two equations together with eq. (\ref{Dnfinal}) then determine
the spectral density for $\nu=1$ shown in figure \ref{rho1n4nu1},
\be
\rho_1^{\nu=1}(x)=S^{\nu=1}_n(x,x)\ .
\ee

We can now perform the limit $a\to0$ as a check. The last term in
eq. (\ref{Snu1}) gives a delta-function, $\delta(x+m)$, the exact
zero-eigenvalue of the $\nu=1$ chGUE shifted to $x=-m$. Using the
result eq. (\ref{rhonu0lim}) and the above results for the limit at
$\nu=0$ we obtain\footnote{From this we might speculate that the second
term in eq. (\ref{Snu1}) leading to $\delta(x+m)$ describes the
distribution of a single eigenvalue at finite-$N$.}
\bea
\lim_{a\to0}\rho^{\nu=1}_1(x)&=& |x|\,\e^{-\frac{x^2}{2}+\frac{m^2}{2}}
\sum_{j=0}^{n-1}L_j\left(\frac{x^2-m^2}{2}\right)^2\Theta(|x|-m)\nn\\
&& -\ |x| \,\e^{-\frac{x^2}{2}+\frac{m^2}{2}}L_n\left(\frac{x^2-m^2}{2}\right)
L^{(1)}_{n-1}\left(\frac{x^2-m^2}{2}\right)\Theta(|x|-m) 
+\delta(x+m).
\eea
This can be seen as follows to be equivalent to the density of the chGUE at
$\nu=1$
\be
\rho_{1,\,\nu=1}^{chGU\!E}(y)= |y|^3
\e^{-y^2/2}\sum_{l=0}^{n-1}\frac{1}{2(l+1)}L_l^{(1)}(y^2/2)^2 \ ,
\label{rhochGUEnu1}
\ee
shifted as in eq. (\ref{shift}). When using further identities \cite{Grad}
\be
XL_l^{(1)}(X)= (l+1)\Big(L_l(X)-L_{l+1}(X) \Big)\ ,\ \ L_l^{(1)}(X)=
L_l(X)+L_{l-1}^{(1)}(X)\ ,
\ee
(and setting $L_{-1}\equiv0$ here), we obtain the desired result, with
$X\equiv\frac12 (x^2-m^2)$, 
\bea
X\sum_{l=0}^{n-1}\frac{1}{l+1}L_l^{(1)}(X)^2 &=&
\sum_{l=0}^{n-1}L_l^{(1)}(X)\Big(L_l(X)-L_{l+1}(X) 
\Big)\nn\\
&=&  \sum_{l=0}^{n-1}\Big(L_l(X)+L_{l-1}^{(1)}(X)\Big)L_l(X) \ -
\sum_{l=0}^{n-1}L_l^{(1)}(X)L_{l+1}(X)\nn\\ 
&=&  \sum_{l=0}^{n-1}L_l(X)^2 \ -\ L_{n-1}^{(1)}(X)L_n(X)\ .
\eea

\sect{The microscopic weakly non-chiral large-N limit}\label{largeN}

The microscopic large-$N$ limit is defined by the following rescaling
of variables 
\be
\hat{x}\equiv \sqrt{2n}\,x\ ,\ \
\hat{a}\equiv \frac12\sqrt{2n}\,a \ .
\label{micro}
\ee
These variables are then to be identified with the variables in WchPT
as given in eq. (\ref{mza-rel}).
Here we send the eigenvalues to zero and $n\to\infty$ such that $\hx$
remains finite, hence we are in the vicinity of the origin. The masses
$m$ and $z$ are rescaled likewise. 
Also the parameter $a$ corresponding to the effect of finite-lattice
spacing in WchPT is sent to zero and $n\to\infty$ such that $\ha$
remains finite. Thus we are in the vicinity of the chGUE 
ensemble which has chiral symmetry. For this reason we call this microscopic
origin limit weakly non-chiral.

We begin with the asymptotic formulas of the partition
function. Because we know the 
($N_f=1$)-flavour partition function for finite-$N$ and general $\nu\geq0$
we  compute its asymptotic first. It reduces to the
asymptotic for the even polynomials $R_{2j}$ at $\nu=0$.
Based on the standard limit \cite{Grad}
\be
\lim_{n \rightarrow \infty}n^{-\nu} L_n^{(\nu)}\Big(\frac{x}{n}\Big)=
x^{-\nu/2}J_\nu(2\sqrt{x})\ ,
\label{Lnlim}
\ee
we obtain the following expression:
\bea
\lim_{n,j\to\infty}\frac{(-)^j}{2^jj!}
\langle\det[x+\Dirac_5]\rangle&=&\lim_{n\to\infty}
(1-a^2)^j
\int_{-\infty}^\infty \frac{ds}{\sqrt{\pi}}\ \e^{-s^2}
(x+2ias-m)^\nu
L_j^{(\nu)}
\left( \frac{(x+2ias)^2-m^2}{2(1-a^2)}\right)\nn\\
&=&\e^{-2t\hat{a}^2}\int_{-\infty}^\infty\frac{ds}{\sqrt{\pi}} \ \e^{-s^2}
t^{\frac{\nu}{2}}\left(\frac{\hat{m}
-(\hat{x}+4is\hat{a})}{\hat{m}+\hat{x}+4is\hat{a}}
\right)^{\frac{\nu}{2}}
I_\nu\Big( \sqrt{t(\hat{m}^2-(\hat{x}+4is\hat{a})^2)}\Big).
\nn\\
\label{vevlim}
\eea
First we have divided out part of the normalisation that cancels with
the norms $r_j$ in the sums inside the kernels. Second we have given the
result for $j=tn$ with $t\in[0,1]$ as we will need these later when
replacing the sum inside the kernel by an integral. Last we have used
that due to the rescaling the prefactor turns into an exponential, 
\be
(1-ja^2/j)^j \to \exp[-ja^2] =\exp[{-2t\hat{a}^2}]\ .
\label{prefac}
\ee
The result eq. (\ref{vevlim}) matches that of the partition function
eq. (\ref{Z1lim}) as it should.

The limiting even polynomial $R_{2j}(x)$ for $\nu=0$ 
is obtained by simply setting $\nu=0$ everywhere in eq. (\ref{vevlim})
(due to parity 
there is no need to switch $x\to -x$ under the determinant):
\bea
\lim_{n,j\to\infty}\frac{(-)^j}{2^jj!}R_{2j}(x)
&=&\e^{-2t\hat{a}^2}\int_{-\infty}^\infty\frac{ds}{\sqrt{\pi}} \e^{-s^2}
I_0\Big( \sqrt{t(\hat{m}^2-(\hat{x}+4is\hat{a})^2)}\Big)\ .
\nn\\
\label{Relim}
\eea
The computation of the limiting odd polynomials $R_{2j+1 }(x)$ for
$\nu=0$ is now a simple consequence. 
Starting from the explicit form eq. (\ref{sOPs}) we obtain
\bea
\lim_{n,j\to\infty}\sqrt{2n}\frac{(-)^j}{2^jj!}R_{2j+1}(x)
&=&\e^{-2t\hat{a}^2}\int_{-\infty}^\infty\frac{ds}{\sqrt{\pi}} \
e^{-s^2}(\hx+4is\ha)
I_0\Big( \sqrt{t(\hat{m}^2-(\hat{x}+4is\hat{a})^2)}\Big)\ ,
\eea
after an integration by parts. As a check the asymptotic of the odd
polynomial is an odd 
function in $\hx$. Note that both prefactors from eq. (\ref{prefac})
will cancel with the limiting norms $r_j$.

\begin{figure}[-h]
\centerline{\epsfig{figure=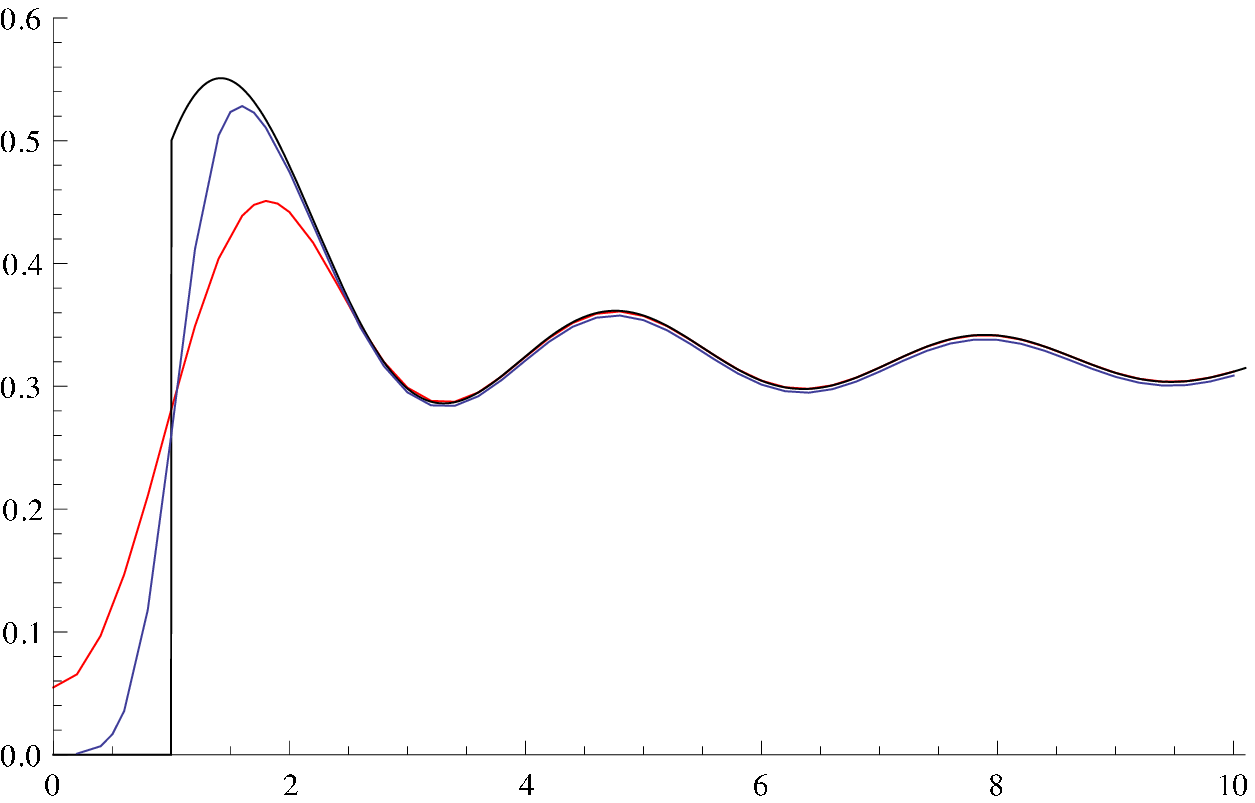,width=15pc}
\epsfig{figure=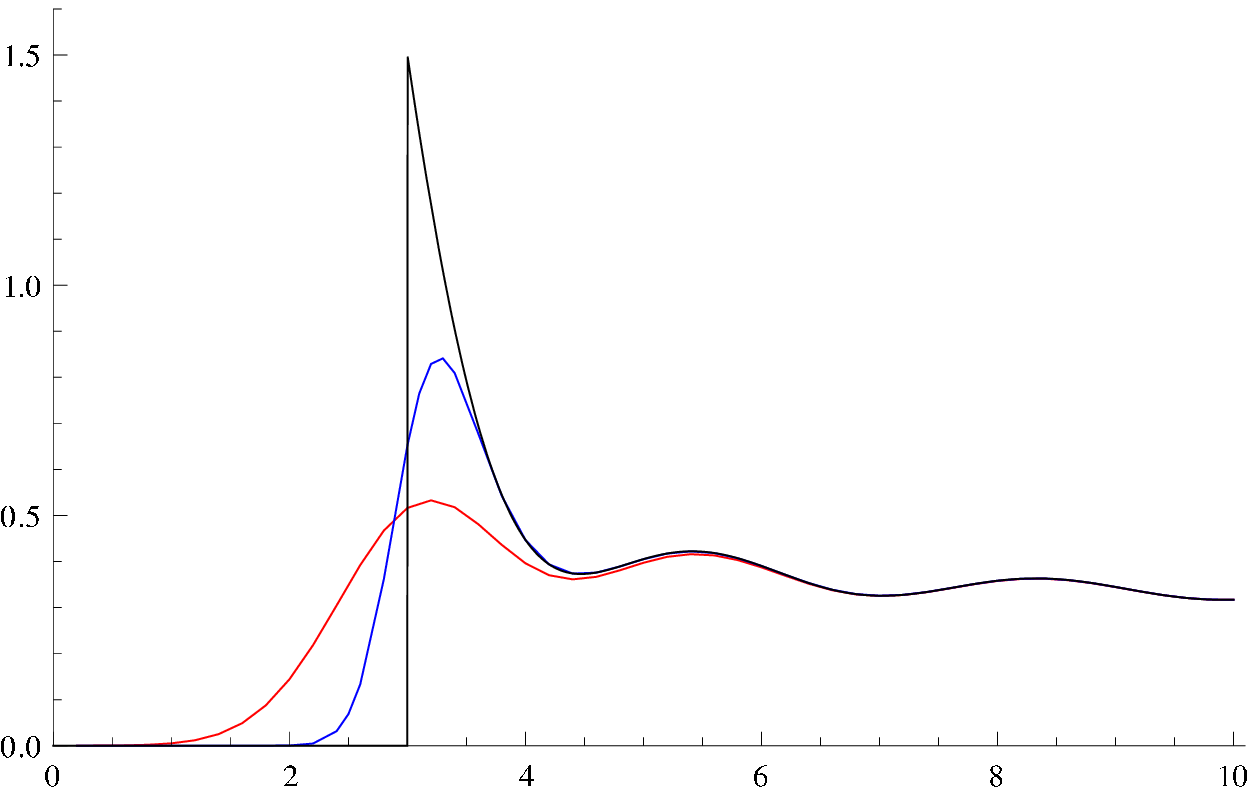,width=15pc}
}
\centerline{\epsfig{figure=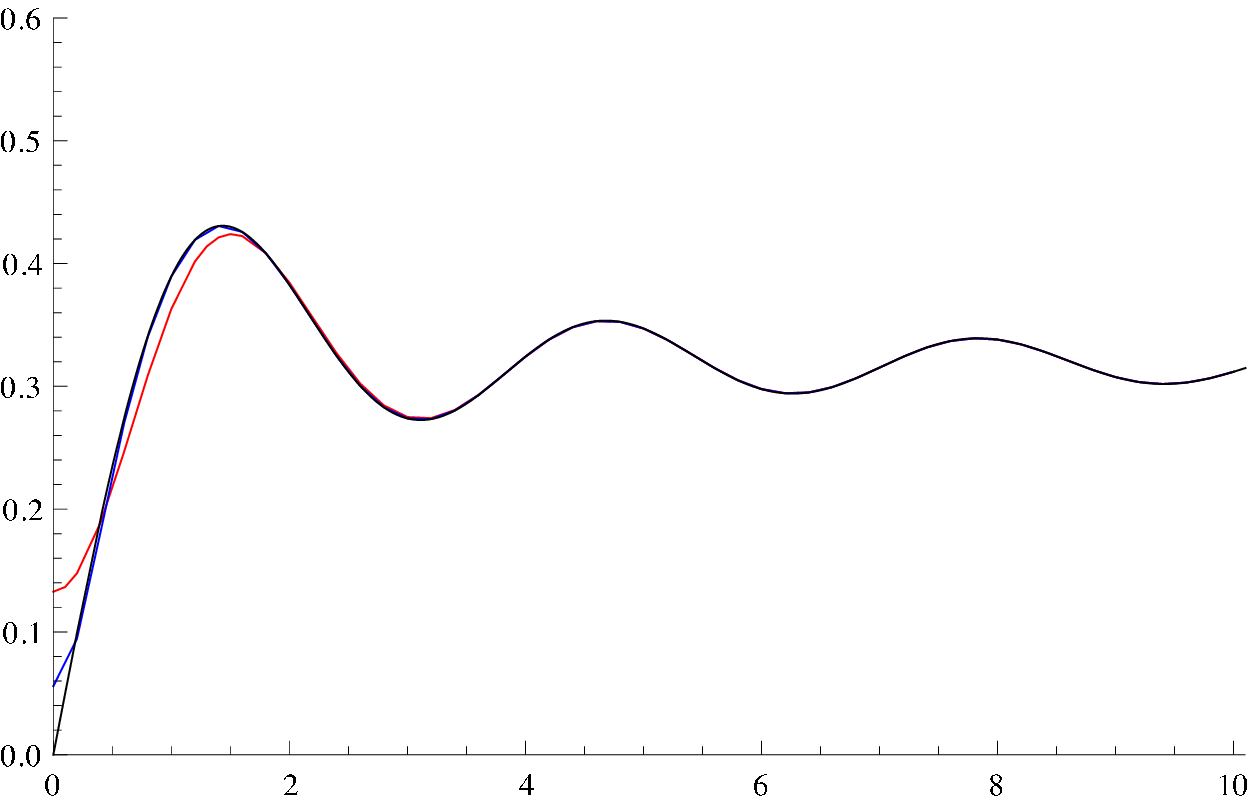,width=15pc}}
\caption{
The microscopic density eq. (\ref{rhomicro})
for $\hm=1$ (top left plots) and $\hm=3$ (top right plots)
at various values of $\ha=0.1$ (middle blue curves) and $\ha=0.25$
(lower red curves).
For comparison we also show the corresponding chGUE density ($\ha=0$) 
eq. (\ref{rhoBes}) shifted as in eq. (\ref{shift}) (top black curves).
For the chGUE-GUE transition at $\hm=0$ (bottom plots) the difference
to $\ha=0.1$ or $\ha=0.25$ is almost indistinguishable, apart from the origin.
}
\label{rhoSplot}
\end{figure}
For the limiting weight $F(x)$ from eq. (\ref{Fdef}) we get a Gaussian
times the error-functions:
\be
\lim_{n\to\infty}F(x)
=\exp\Big[\frac{\hx^2}{32\ha^2}\Big]\left[
\erf\left(\frac{\hx}{4\sqrt{2\ha^2}}+\frac{\hm}{2\sqrt{2\ha^2}}\right)
+\erf\left(\frac{\hx}{4\sqrt{2\ha^2}}-\frac{\hm}{2\sqrt{2\ha^2}}\right)\right].
\ee
Collecting all results we obtain the quenched $\nu=0$ microscopic density
$\rho_S(\hx)\equiv\lim_{n\to\infty}\frac{1}{\sqrt{2n}}\rho_1(x)$:
\bea
\rho_S(\hx)&=&
\int_{-\infty}^\infty
d\hat{y}\frac{\e^{-\frac{(\hx+\hy)^2}{32\ha^2}}}{32\sqrt{2\ha^2\pi}}
\left(\erf\Big(\frac{\hy-\hx+2\hm}{4\sqrt{2\ha^2}}\Big)
+\erf\Big(\frac{\hy-\hx-2\hm}{4\sqrt{2\ha^2}}\Big)
\right)\int_{-\infty}^\infty \frac{ds\,dr}{{\pi}}\e^{-s^2-r^2}\nn\\
&&\times
\int_0^1dt
(\hx-\hy+4i\ha(r-s))
I_0\Big( \sqrt{t(\hat{m}^2-(\hat{y}+4is\hat{a})^2)}\Big)
I_0\Big( \sqrt{t(\hat{m}^2-(\hat{x}+4ir\hat{a})^2)}\Big)
\nn\\
&=&
\int_{-\infty}^\infty
d\hat{y}
\frac{\e^{-\frac{(\hx+\hy)^2}{32\ha^2}}}{16\sqrt{2\ha^2\pi}}
\left(\erf\Big(\frac{\hy-\hx+2\hm}{4\sqrt{2\ha^2}}\Big)
+\erf\Big(\frac{\hy-\hx-2\hm}{4\sqrt{2\ha^2}}\Big)
\right)\int_{-\infty}^\infty \frac{ds\,dr}{{\pi}}\e^{-s^2-r^2}\nn\\
&&\times
\frac{
\sqrt{\hat{m}^2-(\hat{x}+4ir\hat{a})^2}
I_0\Big( \sqrt{\hat{m}^2-(\hat{y}+4is\hat{a})^2}\Big)
I_1\Big( \sqrt{\hat{m}^2-(\hat{x}+4ir\hat{a})^2}\Big)-(\hx\leftrightarrow\hy)
}{\hx+\hy+4i\ha(s+r)}\ .
\nn\\
\label{rhomicro}
\eea
It is shown in figure \ref{rhoSplot}. Here we give both forms, before
and after using the 
Christoffel-Darboux identity which has its large-$N$ correspondence in 
\be
\int_0^1 dt
I_0(X\sqrt{t})I_0(Y\sqrt{t})=\frac{2XI_0(Y)I_1(X)-2YI_0(X)I_1(Y)}{X^2-Y^2}\ .
\ee
Unfortunately we have been unable to check analytically that
eq. (\ref{rhomicro}) is equivalent to the corresponding density
given in \cite{DSV}. There it is given as the discontinuity of the
imaginary part of its resolvent. 

As a check we can again take the limit $\ha\to0$, and we obtain from
the first form in eq. (\ref{rhomicro})
\bea
\lim_{\ha\to0}\rho_S(\hx)&=&\frac{|\hx|}{2}\Theta(|\hx|-\hm)\int_0^1dt
J_0\left(\sqrt{t(\hx^2-\hm^2)}\right)^2\nn\\ 
&=&\frac{|\hx|}{2}\Theta(|\hx|-\hm)\Big[J_0\left(\sqrt{\hx^2-\hm^2}\right)^2+
J_1\left(\sqrt{\hx^2-\hm^2}\right)^2\Big]. 
\eea
It equals the shifted microscopic density of the chGUE at $\nu=0$ after
the shift in eq. (\ref{shift})
\be
\rho_S^{chGU\!E}(\hx)=\frac{|\hx|}{2}\Big( J_0(\hx)^2+J_1(\hx)^2\Big)\ .
\label{rhoBes}
\ee

\begin{figure}[-h]
\centerline{\epsfig{figure=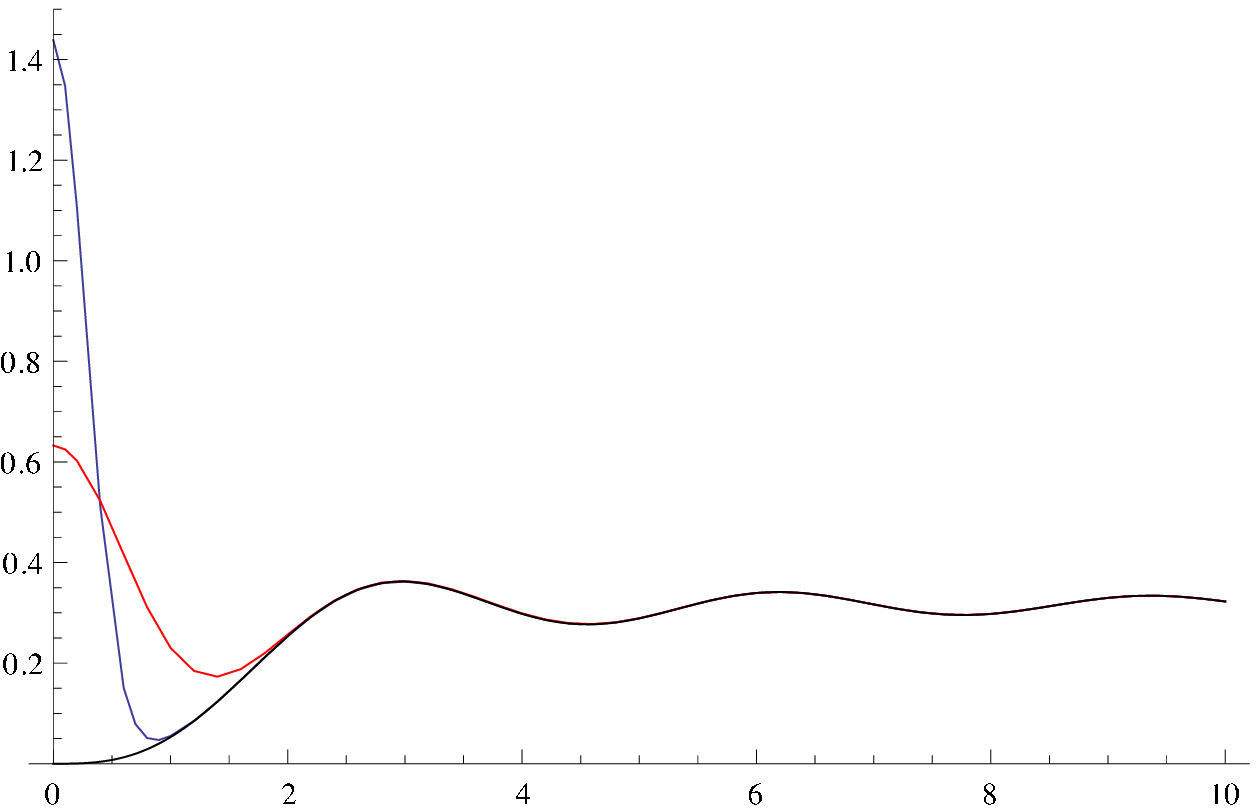,width=15pc}
\epsfig{figure=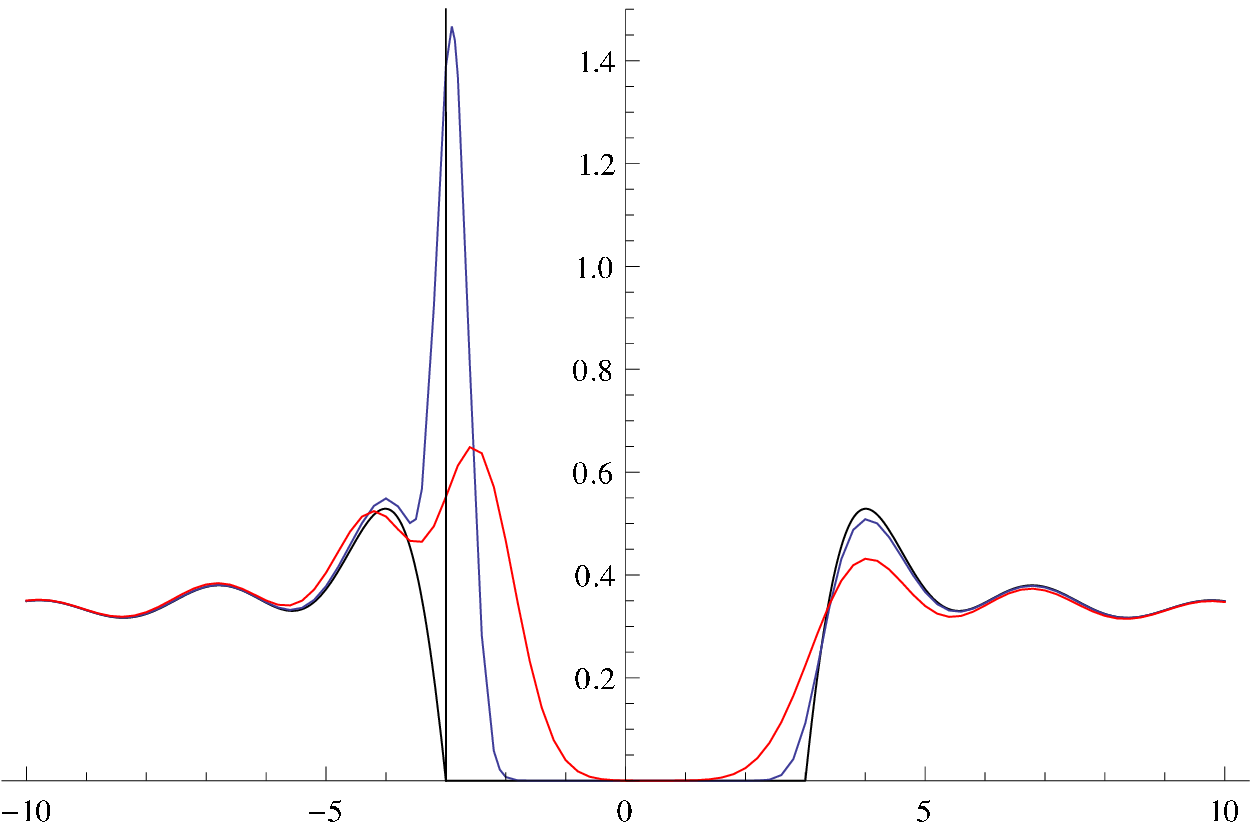,width=15pc}
}
\caption{
The microscopic density eq. (\ref{rhoSnu1})
for $\hm=0$ (left plots) and $\hm=3$ (right plots)
at various values of $\ha=0.3$ (red curves), $\ha=0.1$
(blue curves), and $\ha=0$ (black curves) for comparison.
For the chGUE-GUE transition at $\hm=0$ the density is still
symmetric and we only display $\hx>0$. For $\hm>0$ the delta-function
is shifted to the left, and we can nicely see its increasing broadening
from $\ha=0.1$ to $\ha=0.25$.
}
\label{rhoSnu1plot}
\end{figure}

We now turn to the limiting density for $\nu=1$. Starting from
eqs. (\ref{Dnu1}) and (\ref{Snu1}) it is useful to write it as 
the limiting density at $\nu=0$ which we have already determined, plus a
correction term. Collecting all terms and using the asymptotic
expressions from above we arrive at
\bea
\rho_S^{\nu=1}(\hx)&=& \rho_S(\hx)\ +\ \int_{-\infty}^\infty
d\hat{z}\frac{\e^{-\frac{(\hx+\hz)^2}{32\ha^2}}}{16\ha\pi\sqrt{2\pi}}
\left(\erf\Big(\frac{\hx-\hz+2\hm}{4\sqrt{2\ha^2}}\Big)
+\erf\Big(\frac{\hx-\hz-2\hm}{4\sqrt{2\ha^2}}\Big)
\right)\nn\\
&&\times\left\{\int_{-\infty}^\infty ds\,
\e^{-s^2}I_0\Big( \sqrt{\hat{m}^2-(\hat{x}+4is\hat{a})^2}\Big)
\int_{-\infty}^\infty dr\,\e^{-r^2}{\frac{\hm+\hz+4i\ha
    r}{\sqrt{\hm^2-(\hz+4i\ha r)^2}}}
\right.\nn\\
&&\ \ \ \ \ \ \ \ \times\left.I
_1\Big( \sqrt{\hat{m}^2-(\hat{z}+4ir\hat{a})^2}\Big)
\ -\ (\hx\leftrightarrow\hz)
\right\}\nn\\
&&+\
\frac{1}{4\pi\ha}\e^{-\frac{(\hx+\hm)^2}{16\ha^2}}\int_{-\infty}^\infty
ds\,\e^{-s^2}I_0\Big( \sqrt{\hat{m}^2-(\hat{x}+4is\hat{a})^2}\Big) ,
\label{rhoSnu1}
\eea
which we illustrate in figure \ref{rhoSnu1plot}.

Once more we can perform as a check the limit $\ha\to0$, and we obtain
\bea
\lim_{\ha\to0}\rho_S^{\nu=1}(\hx)&=&
\lim_{\ha\to0}\rho_S(\hx)-\frac{|\hx|}{\sqrt{\hx^2-\hm^2}}
J_0\left(\sqrt{\hx^2-\hm^2}\right)
J_1\left(\sqrt{\hx^2-\hm^2}\right)\Theta(|\hx|-\hm)
\ +\ \delta(\hx+\hm)\ .\nn\\
\label{rhonu1lim}
\eea
The matching with the density of the chGUE at $\nu=1$,
\be
\rho_{S,\,\nu=1}^{chGU\!E}(\hx)=\frac{|\hx|}{2}
\Big( J_1(\hx)^2-J_0(\hx)J_2(\hx)\Big)\ ,
\label{rhoBesnu1}
\ee
shifted as in eq. (\ref{shift}) can be seen as follows:
\bea
\rho_{S,\,\nu=1}^{chGU\!E}(X)-\rho_{S}^{chGU\!E}(X)&=&
-\frac{X}{2}J_0(X)(J_2(X)+J_0(X))
\nn\\
&=&-J_0(X)J_1(X)\ .
\eea
Here we used a Bessel identity and the abbreviation $X=\sqrt{\hx^2-\hm^2}$. The
delta-function in eq. (\ref{rhonu1lim}) 
is just the exact zero-eigenvalue at $\hx=-\hm$, originating from the
last term in eq. (\ref{rhoSnu1}) which is non-Gaussian.

\sect{Conclusions}\label{conc}

In this paper we have introduced a parameter dependent Gaussian random
two-matrix model that interpolates between the chGUE ($a=0$) and the GUE
($a=1$) symmetry class (possibly shifted by a constant matrix). 
At the same time this model describes the
spectral properties of the Hermitian Wilson Dirac operator 
when we take the microscopic weakly non-chiral limit at the origin. Here the
rescaled parameter $\ha\sim a \sqrt{n}$ is interpreted as the effect
of finite lattice
spacing. It is properly incorporated in a continuum effective field
theory as
Wilson chiral Perturbation theory in the epsilon regime to order
${\cal O}(\ha^2)$.  

We have completely determined all spectral $k$-point density
correlation functions when starting from a chGUE of size
$(2n+\nu)$ with $\nu=0$ or $\nu=1$ exact zero modes, 
both for finite $n$ and in the microscopic large-$n$ origin
limit at weak non-chirality. 
They are given in terms of the Pfaffian of a  $2\times 2$
matrix kernel, containing three individual kernels as building
blocks. Each kernel is expressed as a sum over skew-orthogonal
polynomials (and their integral transforms) depending on $\nu=0$ or 1, 
which we explicitly constructed 
in terms of Gaussian integrals over Laguerre polynomials. 
In the large-$n$ limit these 3 respective kernels contain in their final form a
fixed number of 2 up to 4 Gaussian integrals over modified Bessel
functions, respectively. The Pfaffian of the matrix kernel is thus
easily evaluated for any $k$-point function. 

Let us compare to what is known in the literature. In \cite{DSV,ADSV}
the supersymmetric method was used to compute the microscopic spectral
density ($k=1$) directly from the chiral Lagrangian for arbitrary
$\nu$, the so-called index. In \cite{ADSVLatNf} this was extended to
include one ($N_f=1$) extra quark determinant. This method rapidly
becomes cumbersome when adding more flavours $N_f>1$ or increasing
$k$, as the number of integrations increases with the dimension of the
auxiliary supermatrix from the Hubbard-Stratonovich transformation. 
This problem was partly circumvented in \cite{SV2011} by using the graded
supereigenvalue method, where explicit examples where provided for the
density $k=1$ up to $N_f=2$, and for the two-point density $k=2$
quenched. However, also here it is difficult to establish the Pfaffian
structure of the $k$-point functions, and thus the number of integrals
grows with $k$ and $N_f$ for the $k$-point density correlation
function (for recent progress see however \cite{KV}). 

It is an open problem how to include both $N_f>0$ and $\nu>1$ in our
skew-orthogonal polynomial method, in particular the latter, as the
joint probability distribution of eigenvalues which we computed here
for the general $N_f$ and $\nu$ case is no longer a simple
Pfaffian of a single block matrix for $\nu>0$. The reason for attempting this 
problem in the future is
not only an independent confirmation for the aforementioned analytical results
which are mainly for $k=1$. The extension to arbitrary $k$ is crucial
if we want to get a hand on individual eigenvalue distributions 
at least in a perturbative expansion (see e.g. \cite{AD03}),
apart from the conceptual aspect of integrability of our two-matrix model. 
Furthermore, in all works \cite{DSV,ADSV,ADSVLatNf,SV2011} also the
density of real eigenvalues of the complex non-Hermitian Wilson Dirac
operator was considered. The extension to this operator is another
challenge to our method. 
\\

{\bf Acknowledgments:} It is a pleasure to thank P. Damgaard,
M. Kieburg, K. Splittorff, and J. Verbaarschot for collaborations on
related topics and many discussions. The Niels Bohr Institute
and Niels Bohr International Academy are thanked (G.A.) for
hospitality and partial financial support when this work was
initiated. 
This work was partially supported by the Japan Society 
for the Promotion of Science (KAKENHI 20540372) (T.N.).

\begin{appendix}

\sect{Equivalence to the partition function of WchPT}\label{equiv}

In this appendix we show the equivalence of our matrix model in the
microscopic limit to WchPT in the epsilon regime, as it was displayed
in eq. (\ref{match}). For non-degenerate parameters we have 
\bea
{\cal Z}_{\nu}^{(N_f)}
&\sim&\int dAdBdWd\Omega
\prod_{f=1}^{N_f}\det\left[
\begin{array}{cc}
(z_f+m_f)\eins_n+A & W+\Omega\\
W^\dag+\Omega^\dag & (z_f-m_f)\eins_{n+\nu}+B\\
\end{array}
\right]\e^{-\frac{1}{2(1-a^2)}\Tr(WW^\dag)}
\nn\\
&&\times \e^{-\frac{1}{4a^2}\Tr(A^2+B^2+2\Omega\Omega^\dag)}\ .
\eea
Defining $z_f^\pm\equiv z_f\pm m_f$ we now express the determinants
as integrals over $N_f$ sets of complex Grassmann
variables $\eta_i^f$, $i=1,\ldots,n$ and $\psi_b^f$, $b=1,\ldots,n+\nu$,
\bea
\prod_{f=1}^{N_f}\det[\ldots]&=& \int d^2\eta d^2\psi
\e^{\eta_i^{*\,f}(z^+_f\delta_{ij}+A_{ij})\eta_j^f
+\psi^{*\,f}_b(W_{bj}^\dag+\Omega_{bj}^\dag)\eta_i^f
+\eta_i^{*\,f}(W_{ib}+\Omega_{ib})\psi_b^f
+\psi^{*\,f}_c(z_f^-\delta_{cb}+B_{cb})\psi_b^f
}\ ,\ \ \ \ 
\eea
using summation conventions in all indices.
A completion of the squares and integration of the Gaussian matrices yields
\bea
{\cal Z}_{\nu}^{(N_f)}
&\sim&
\int d^2\eta d^2\psi\
e^{z^+_f\eta^{*\,f}_i\eta_i^f + z^-_f \psi^{*\,f}_b\psi_b^f-a^2
\left((\eta^{*\,f}_i\eta_i^f)^2+(\psi^{*\,f}_b\psi_b^f)^2\right)
-2\psi^{*\,f}_b\psi_b^f\eta^{*\,g}_i\eta_i^g}\nn\\
&=&\int dQ_1dQ_2dT \det[Z^+-i\sqrt{2}(aQ_1+T^\dag)]^n
\det[Z^- -i\sqrt{2}(aQ_2+T)]^{n+\nu}
\e^{-\Tr (TT^\dag+Q_1^2+Q_2^2)}\ .\nn\\
&&
\label{ZINfexact}
\eea
Here we have introduced two Hermitian matrices $Q_{1,2}$ and one
complex non-Hermitian matrix $T$ of size $N_f\times N_f$ 
to do the Hubbard-Stratonovich transformation. Integrating out all
Grassmann variables yields an exact expression containing the diagonal
matrices $Z^\pm$ with elements $z_f^\pm$ (up to some constant prefactors that we
dropped). If we parametrise $T=UR$ with $U$ unitary and $R$ positive
definite Hermitian the saddle point is taken at
$R\sim\eins_{N_f}\sqrt{n}$ when we employ the scaling of masses and
$a$ indicated in eq. (\ref{mza-rel}). We obtain 
\bea
\lim_{n\to\infty}{\cal Z}_{\nu}^{(N_f)}&\sim& \int dQ_1dQ_2\int dU \det[U]^\nu
\e^{\frac{i}{2}\Tr(\hat{Z}^+U+\hat{Z}^-U^\dag)+\sqrt{2\ha^2}\Tr(Q_1U+Q_2U^\dag)
-\frac12\Tr(Q_1^2+Q_2^2)}\nn\\
&\sim& \int dU \det[U]^\nu
\exp\Big[\frac{1}{2}\Tr(\hat{M}(U+U^\dag))
+\frac12\Tr(\hat{Z}(U-U^\dag))-\ha^2\Tr(U^2+U^{\dag\,2})\Big]\
,\ \ \ \ 
\label{ZPTNF}
\eea
after integrating out the matrices $Q_{1,2}$ and performing a further
rotation $U\to iU$, $U^\dag\to-iU^\dag$. Eq. (\ref{ZPTNF}) is the
partition function of WchPT in the epsilon regime for non-degenerate
masses, hence we have completed our claimed proof.

\subsection{Alternative representation of $N_f=1$ WchPT}

Finally we would like to make the equivalence between the
$N_f=1$ flavour partition function for the
large-$N$ limit of our even polynomials $R_{2n}(x)$ explicit. The
reason why we present this check in detail is that we require a
slightly different form from the one given in \cite{DSV} eq. (13), or
in \cite{ADSV} eq. (151) (for $\hz=0$ there):
\bea
{Z}_\nu^{(N_f=1)}(\hm,\hz,\ha) &=& 
\e^{-2\ha^2}\int_{-\pi}^\pi \frac{d\theta}{2\pi} \e^{i\theta\nu}
\exp[\hm\cos(\theta)+i\hz\sin(\theta)+4\ha^2\sin^2(\theta)]\nn\\
&=& \e^{-2\ha^2}\int_{-\pi}^\pi \frac{d\theta}{2\pi} \e^{i\theta\nu}
\int_{-\infty}^\infty \frac{dx}{\sqrt{\pi}} \e^{-x^2}
\exp[\hm\cos(\theta)+i\hz\sin(\theta)-4x\ha\sin(\theta)]\ ,\ \ \ \ 
\eea
where after using a trigonometric identity we have linearised the
sine-squared term instead. 
We can now use the following identity \cite{BT} for generic matrices $A,B$
\be
\int_{U(N_f)}\det[U]^\nu\exp\left[\frac12 \Tr(AU+BU^\dag)\right]
\ =\
2^{N_f(N_f-1)/2}\prod_{j=1}^{N_f-1}j!\
\left(\frac{\det[B]}{\det[A]}\right)^{\nu/2}
\frac{\det[\mu_i^{j-1}I_{\nu+j-1}(\mu_i)]}
{\Delta_{N_f}(\{\mu^2\})}\ ,
\ee
with $\mu_i^2$ being the eigenvalues of the product matrix
$AB$. Specifying to $U=e^{i\theta}\in U(1)$ and identifying
$A=\hm+\hz+4ix\ha$, $B=\hm-(\hz+4ix\ha)$ we arrive at
\be
{Z}_\nu^{(N_f=1)}(\hm,\hz,\ha)  
= \e^{-2\ha^2}\int_{-\infty}^\infty \frac{dx}{\sqrt{\pi}} \e^{-x^2}
\left(\frac{\hm-(\hz+4ix\ha)}{\hm+\hz+4ix\ha}\right)^{\nu/2}
I_\nu\Big(\sqrt{\hm^2-(\hz+4ix\ha)^2} 
\Big)\ ,
\label{Z1lim}
\ee
which agrees perfectly with eq. (\ref{vevlim}), see also eq. (99) in
\cite{SV2011}.

\sect{Consistency checks of the partition function and
  factorisation}\label{Zjpdfgen} 

In this appendix we perform a few consistency checks of the eigenvalue
representation of the partition function eq. (\ref{ZDevPf}). For
simplicity we will set $m=0$ and $N_f=0$ here. The antisymmetric weight
eq. (\ref{Fdef}) thus reduces to 
\be
F(x)|_{m=0}=2\ \e^{\frac{x^2(1-a^2)}{8a^2}}
\erf\Big(\frac{x\sqrt{1-a^2}}{2\sqrt{2a^2}}\Big)\ .
\label{Fm0}
\ee

\subsection{The GUE limit $a\to 1$ for $\nu=0$}

On the level of
the distribution eq. (\ref{P12}) it is clear that the matrix $W$ gets
suppressed for $a\to1$. After reducing to eigenvalues the same result
should come out. 
Here we will look at the simplest case only, that is $\nu=0$.
We can insert the simplification  eq. (\ref{Fm0})
into eq. (\ref{Zjpdf}) at $\nu=0$. Noting that in
the limit $a\to1$ the exponential factor in eq. (\ref{Fm0}) reduces to unity,
we can use the following result of \cite{PandeyMehta} (see
end of section 3 in the first paper) 
for the limit of a Pfaffian of an error-function,
\be
\lim_{a\to1} \Pf_{1\leq i,j\leq 2n}\left[
\erf\left( (d_{i}-d_{j}) \sqrt{\frac{(1-a^2)}{8a^2}}\right)
\right]
\sim\ \Delta_{2n}(\{d\})\ .
\ee
This provides the second Vandermonde determinant in eq. (\ref{Zjpdf}) to
constitute the GUE, and we arrive at
\be
\lim_{a\to1}{\cal Z}_{\nu=0}^{(N_f=0)}|_{m=0} \sim
\int_{-\infty}^\infty\prod_{j=1}^{N}
dd_j\,d_j^{N_f}\exp\Big[{-\frac{d_j^2}{4}}\Big]
\ \Delta_{N}(\{d\})^2\ .
\label{ZGUE}
\ee

\subsection{The chGUE limit $a\to0$ and factorisation}

Here we would like to perform a more detailed check. As it was shown
in \cite{ADSV} 
in the large-$N$ limit
the spectral density of the Hermitian Wilson Dirac operator behaves
like the density of a 
{\it finite} GUE-matrix of size $\nu\times\nu$ plus the microscopic
chGUE density. We therefore expect
that our general partition function will factorise in the limit
$a\to0$ into a finite GUE times a second chGUE partition function of the
remaining $2n$ eigenvalues. 

In order to show that we will first simplify ${\cal Z}_\nu^{(0)}$
further before using the de Bruijn integration theorem. Taking
eq. (\ref{Zev}) as a starting point we can use the fact that the jpdf
integrated over all variables 
is totally symmetric under exchange or relabeling of the
$d_j$ \footnote{Using this symmetry our manipulations from now on will
  not be valid 
in general for correlation functions.}.
We can split the variables into two sets, denoted by
$t_{j=1,\ldots,\nu}\equiv d_{j=1,\ldots,\nu}$, and
$s_{j=1,\ldots,2n}\equiv d_{j=\nu+1,\ldots,2n+\nu}$. If we do a Laplace
expansion of the determinant into blocks of sizes $2n\times 2n$ and
$\nu\times\nu$, we can split off a Vandermonde type determinant in the
variables $t_j$, after appropriately relabeling all other permutations:
\bea
{\cal Z}_\nu^{(0)} &\sim&
\int_{-\infty}^\infty\prod_{j=1}^{\nu}
dt_j\,\e^{-\frac{t_j^2}{4a^2}}
\Delta_{\nu}(\{t\})
\int_{-\infty}^\infty\prod_{j=1}^{2n}
ds_j\,\e^{-\frac{s_j^2}{4a^2}}\prod_{i=1}^{\nu}(s_j-t_i)
\ \Delta_{2n}(\{s\})
\label{ZprePf}\\
&&\times
\int_0^\infty\prod_{b=1}^{n} du_b
\e^{\frac{- u_b^2}{2a^2(1-a^2)}}
\left|
\begin{array}{cc}
\e^{\frac{s_1u_1}{2a^2}}\ldots
\e^{\frac{s_1u_n}{2a^2}}&\e^{-\frac{s_1u_1}{2a^2}}\ldots
\e^{-\frac{s_1u_n}{2a^2}}\\ 
\cdots          & \cdots\\
\e^{\frac{s_{2n}u_1}{2a^2}}\ldots\e^{\frac{s_{2n}u_n}{2a^2}}
&\e^{-\frac{s_{2n}u_1}{2a^2}}\ldots\e^{-\frac{s_{2n}u_n}{2a^2}}\\  
\end{array}\right|
\left|
\begin{array}{cccc}
1&{t_1}&\cdots&t_1^{\nu-1}\\
\cdots&\cdots          &\cdots& \cdots\\
1&t_{\nu}&\cdots &t_{\nu}^{\nu-1}
\end{array}\right|\!\!.\nn
\eea
We can now perform the integrations over $u_{1},\ldots,u_n$
with the help of the standard de Bruijn integral formula eq. (\ref{gendeB}).
To make it applicable we can simply multiply the weight factors into the
determinant, and we obtain the following result:
\bea
{\cal Z}_\nu^{(0)}\! &\sim&\!
\int_{-\infty}^\infty\prod_{j=1}^{\nu}
dt_j\,\e^{-\frac{t_j^2}{4a^2}}
\Delta_{\nu}(\{t\})^2\!
\int_{-\infty}^\infty\prod_{j=1}^{2n}
ds_j\,\e^{-\frac{s_j^2}{4a^2}}\prod_{i=1}^{\nu}(s_j-t_i)
\Delta_{2n}(\{s\})
\Pf_{1\leq i,j\leq 2n}\left[F(s_i-s_j)
\right]\!.\ \ \ \ \ \ \ 
\label{Zxt}
\eea
In eq. (\ref{Zxt}) we can make further use of the symmetry under
relabeling of the 
variables $s_j$ and the anti-symmetry of the Vandermonde determinant
to write the Pfaffian as the following product: 
\bea
{\cal Z}_\nu^{(0)} &\sim&
\int_{-\infty}^\infty\prod_{j=1}^{\nu}
dt_j\,\e^{-\frac{t_j^2}{4a^2}}
\Delta_{\nu}(\{t\})^2\!
\int_{-\infty}^\infty\prod_{j=1}^{2n}
ds_j\,\e^{-\frac{s_j^2}{4a^2}}\prod_{i=1}^{\nu}(s_j-t_i)
\ \Delta_{2n}(\{s\})
\prod_{k=1}^n F(s_{2k}-s_{2k-1}).\ \ \ \ \ \ 
\label{ZDevm=0}
\eea
Now consider the exponents of two consecutive variables, say $s_1$ and
$s_2$, from the Gaussian and the exponential part of $F$: 
\be
\exp\Big[{-\frac{s_1^2+s_2^2}{4a^2}}+
\frac{(s_{2}-s_{1})^2(1-a^2)}{8a^2}\Big]
=\exp\Big[-\frac{(s_1+s_2)^2}{8a^2}-\frac{(s_1-s_2)^2}{8}\Big].
\ee
In the limit $a\to0$ the first term on the right hand side will give a
delta-function, whereas the second Gaussian term remains:
\be
\lim_{a\to0}\sqrt{8a^2\pi}\exp\Big[-\frac{(s_1+s_2)^2}{8a^2}
-\frac{(s_1-s_2)^2}{8}\Big] = \delta(s_1+s_2)\ \e^{-s_1^2/2}\ .
\ee
In the same limit the error-function in eq. (\ref{Fm0}) becomes the
sign function (see e.g. \cite{PandeyMehta}), 
\be
\lim_{a\to0}\erf\Big[ (s_2-s_1)
\sqrt{\frac{(1-a^2)}{8a^2}}\Big]=\sign[s_2-s_1]\ .
\label{limerf}
\ee
We are now ready to take the limit $a\to0$ of the partition function
eq. (\ref{ZDevm=0}). We first formally
change variables $r_j=t_j/a$ 
to isolate the $a$-dependence, and then
take the limit only on the remaining $a$-dependence, that is the
variables $r_j$, 
\be
\lim_{a\to0}{\cal Z}_\nu^{(0)}
\sim\int_{-\infty}^\infty \prod_{j=1}^{\nu} dr_j \, 
\e^{-\frac{r_j^2}{4}}\Delta_{\nu}(\{ r \})^2 
\int_{-\infty}^\infty\prod_{j=1}^{n} ds_j \, 
s_j \sign[s_j] \exp\Big[{-\frac{s_j^2}{2}}\Big]
\prod_{i=1}^{\nu}(s_j^2 - a^2 r_i^2) \ 
\Delta_{n}(\{s^2\})^2.
\label{ZDeva0}
\ee
We have used the fact that for $s_{2k} \to - s_{2k-1}$ we obtain 
$\Delta_{2n}(\{s\}) \to \prod_{i=1}^n 2 s_i \ \Delta_n(\{s^2\})^2$.
The variables $r_j$ thus act as mass terms to the $s$-variables, 
but to leading order these masses can be neglected, leading to 
a complete decoupling of variables.
Our final result is thus reading
\be
\int_{-\infty}^\infty\prod_{j=1}^{\nu} dr_j \, 
\exp\Big[{-\frac{r_j^2}{4}}\Big]
\Delta_{\nu}(\{r\})^2
\int_{-\infty}^\infty\prod_{j=1}^{n}
ds_j\,|s_j|^{2\nu+1}\exp\Big[{-\frac{s_j^2}{2}}\Big]
\ \Delta_{n}(\{s^2\})^2,
\label{Zfactor}
\ee
where the $\nu$ would-be zero-eigenvalues at $a=0$ have decoupled into
a GUE of size $\nu$, 
times the standard massless chGUE with exactly $\nu$ zero eigenvalues.
However, we should remember that due to rescaling
the original variables $t_j$ are all of the order ${\cal O}(a)$.

It is clear how to this order of approximation the microscopic
spectral density would look like: it is the superposition of a
finite-$\nu$ semi-circle from the GUE and 
the Bessel spectral density with $\nu$ zero-eigenvalues from the
chGUE. This confirms the analysis of \cite{ADSV} on the level of the
partition function, to that order.

\sect{Skew-orthogonal polynomials}\label{SOP-C}

In this appendix we compute the skew-orthogonal polynomials given in
eqs. (\ref{sOPs}) for $\nu=0$.
We first recall that the expectation value of a single characteristic
polynomial (or equivalently the $N_f=1$ partition function) gives the
even polynomials $R_{2j}$ where we follow the argument of
\cite{AKP}. This expectation value is
computed using the supersymmetric method in the next subsection
\ref{Reven}. Based on that the odd polynomials $R_{2j+1}$ then follow
as shown in \ref{Rodd}.

Starting from the quenched partition function eq. (\ref{Zjpdf}) for $\nu=0$
\be
{\cal Z}_0^{(0)}
=
\int_{-\infty}^\infty\prod_{j=1}^{2n}
ds_j w(s_j)
\ \Delta_{2n}(\{s\})\ \Pf_{1\leq i,j\leq 2n}[F(s_i-s_j)]\ ,
\ee
we can write the expectation value of a single determinant as follows:
\bea
\langle \det[z-\Dirac_5]\rangle_{2n}
&=&\frac{1}{{\cal Z}_0^{(0)}} \int_{-\infty}^\infty\prod_{j=1}^{2n}
ds_j w(s_j) (z-s_j)
\ \Delta_{2n}(\{s\})\ \Pf_{1\leq i,j\leq 2n}[F(s_i-s_j)]
\nn\\
&=&\frac{1}{{\cal Z}_0^{(0)}}
 \frac{(2n)!}{2^nn!}\int_{-\infty}^\infty\prod_{j=1}^{N}
ds_j w(s_j)
\ \Delta_{2n+1}(\{s\},z)\ \prod_{l=1}^nF(s_{2l}-s_{2l-1}).
\eea
Here we have used the fact that both the Vandermonde determinant and
the Pfaffian are antisymmetric 
in order to simplify their product (times the symmetric integral over
all eigenvalues), cf. \cite{Mehta} Appendix 25. Furthermore we have
included the determinant into a larger Vandermonde.
Writing the Vandermonde determinant in terms of even and odd monic
polynomials $\Phi_{2i}$ and $\Psi_{2i+1}$ respectively,
\be
\Delta_{2n+1}(\{s\},z)
=\det_{i=0,\ldots,n-1;j=1,\ldots,2n+1}\Big[\{\Phi_{2i}(s_j),
\Psi_{2i+1}(s_j)\}\ \Phi_{2n}(s_j) \Big]\ ,
\ee
and relabeling $z=s_{2n+1}$, we have from the generalised de Bruijn
formula (cf. \cite{AKP} eq. (4.9), or \cite{KG})
\be
\langle \det[z-\Dirac_5]\rangle_{2n}=
\frac{n!}{{\cal Z}_0^{(0)}}\ \Pf_{0\leq i,j\leq 2n}
\left[
\begin{array}{lll}
\langle \varphi_i,\varphi_j\rangle & \ldots& \varphi_0(s_{2n+1})\\
\ldots &&\cdots\\
&& \varphi_{2n}(s_{2n+1})\\
-\varphi_0(s_{2n+1})\ldots & -\varphi_{2n}(s_{2n+1})&0\\
\end{array}
\right],
\label{preskew}
\ee
for the set of all monic polynomials $\varphi_{i}=\{
\Phi_0, \Psi_1, \Phi_2, \Psi_3, \cdots, \Phi_{2 n}
\}$, with the scalar product defined in
eq. (\ref{skew}). Choosing polynomials $\varphi_j=R_j$ to satisfy the
skew-orthogonality eq. (\ref{skewdef}) we obtained the desired
relation
\be
R_{2n}(z)=\langle \det[z-\Dirac_5]\rangle_{2n}= \langle
\det[z+\Dirac_5]\rangle_{2n}\ ,
\label{Revenvev}
\ee
which obviously has parity.

\subsection{Expectation value of a single characteristic
  polynomial}\label{Reven} 

We now compute the expectation value of a single characteristic
polynomial eq. (\ref{Revenvev})
using the supersymmetric method for finite-$N$. As we have shown already this
yields the subset of the even skew-orthogonal polynomials $R_{2n}(x)$
from which the odd polynomials are constructed. At the same time this
also gives the one flavour partition function, which we can compare to
the known result from Wilson chiral perturbation theory \cite{DSV} in the
large-$N$ limit as a consistency check.

The calculation we present here for $\nu\geq0$ is slightly more
general than needed in eq. (\ref{sOPs}). Using the definitions
(\ref{D5Idef}) and (\ref{Hdef}) for the Hermitian matrices
$A$ and $B$ of sizes $n\times n$ and $(n+\nu)\times(n+\nu)$ as well as
matrices $W,\Omega$ of sizes $n\times(n+\nu)$ with weights
eqs. (\ref{P12}) we have for $N=2n+\nu$
\bea
\langle \det[z+\Dirac_5]\rangle_N
&=&\frac{1}{{\cal Z}_0^{(0)}}\int dAdBdWd\Omega
\det\!\left[
\begin{array}{cc}
(z+m)\eins_n+A & W+\Omega\\
W^\dag+\Omega^\dag & (z-m)\eins_{n+\nu}+B\\
\end{array}
\right]\e^{-\frac{1}{2(1-a^2)}\Tr(WW^\dag)}
\nn\\
&&\times \e^{-\frac{1}{4a^2}\Tr(A^2+B^2+2\Omega\Omega^\dag)}
\eea
Expressing the determinant through integrals over complex Grassmann
variables $\eta_i$, $i=1,\ldots,n$ and $\psi_b$, $b=1,\ldots,n+\nu$,
\bea
\det[\ldots]&=& \int d^2\eta d^2\psi
\e^{\eta_i^*((z+m)\delta_{ij}+A_{ij})\eta_j 
+\psi^*_b(W_{bj}^\dag+\Omega_{bj}^\dag)\eta_i
+\eta_i^*(W_{ib}+\Omega_{ib})\psi_b+\psi^*_c((z-m)\delta_{cb}+B_{cb})\psi_b
}\ ,
\eea
we obtain the following result after performing the Gaussian integrals
over the matrices $A,B,W,\Omega$:
\bea
\langle \det[z+\Dirac_5]\rangle_N&\sim&
\int d^2\eta d^2\psi\
\e^{(z+m)\eta^*_i\eta_i + (z-m) 
\psi^*_b\psi_b-a^2\left((\eta^*_i\eta_i)^2+(\psi^*_b\psi_b)^2\right)
-2\psi^*_b\psi_b\eta^*_i\eta_i}\nn\\
&=& \int d^2\eta d^2\psi\int_{-\infty}^\infty 
dt\int_{\mathbb C} \frac{d^2u}{\pi}\e^{-t^2-|u|^2}
\exp[(z+m-2iat)\eta^*_j\eta_j +(z+m-2iat) \psi^*_b\psi_b]
\nn\\
&&\times \exp[-i\sqrt{2(1-a^2)}(u\eta^*_j\eta_j + u^*\psi^*_b\psi_b)]\nn\\
&=&\int_{-\infty}^\infty dt\int_{\mathbb C}
\frac{d^2u}{\pi}\e^{-t^2-|u|^2}\Big(z+m+2iat+i\sqrt{2(1-a^2)}u\Big)^n\nn\\
&&\times\Big(z-m+2iat+i\sqrt{2(1-a^2)}u^*\Big)^{n+\nu}\ .
\eea
In the second step we have linearised upon using one real and one
complex variable in our Hubbard-Stratonovich transformation.
As a last step we use the following integral representation for
generalised Laguerre polynomials $L_n^{\nu}(x)$ \cite{APSo}
\be
\int_{\mathbb C}
\frac{d^2u}{\pi}\e^{-|u|^2}(iu+\gamma)^n(iu^*+\la)^{n+\nu}=(-)^n\la^\nu
L_n^{(\nu)}(\la\gamma)\ ,
\ee
to obtain
\bea
\langle \det[z+\Dirac_5]\rangle_N&=& 
\frac{n!\,(2(1-a^2))^{n+\frac{\nu}{2}}}{(-)^n\sqrt{\pi}}
\int_{-\infty}^\infty dt\ \e^{-t^2} \Big(
\frac{z+2iat-m}{\sqrt{2(1-a^2)}}\Big)^\nu
L_n^{(\nu)}\left( \frac{(z+2iat)^2-m^2}{2(1-a^2)}\right)\!.\ \ \ \ \ \
\label{Reresult}
\eea
Here we have put the correct normalisation coefficient that can be
read off from $\langle \det[z+\Dirac_5]\rangle_N=z^{N}+{\cal
  O}(z^{N-1})$ for large $z$. For $\nu=0$ this reduces to the result
given in eq. (\ref{sOPs}). 
It can be easily seen that in the microscopic limit eq. (\ref{micro})
using eq. (\ref{Lnlim}) this reduces to the $N_f=1$ flavour partition
function eq. (\ref{Z1lim})

\subsection{The odd polynomials $R_{2j+1}(x)$}\label{Rodd}

In this subsection we construct the odd set of skew orthogonal
polynomials directly from the even ones by differentiation.
Instead we could have computed the $R_{2j+1}(x)$ again from the
supersymmetric method as in the previous subsection, using the
relation \cite{AKP} $R_{2n+1}(x)=\langle
\det[x-\Dirac_5](x+\Tr\Dirac_5)\rangle_{2n}$.

Based on our result above eq. (\ref{Reresult}) for $\nu=0$
$R_{2n}(x)=\langle \det[x-\Dirac_5]\rangle_{2n}$
we define
\bea
R_{2j+1}(x)&\equiv&-2a^2w(x)^{-1}
\frac{d}{dx}\Big[
w(x)R_{2j}(x)\Big]\ ,\label{Rodddef}\\
w(x)&\equiv&\exp[-x^2/(4a^2)]\ .
\eea
In the following we will show by direct computation that these satisfy
the skew-orthogonality relations eq. (\ref{skewdef}) with respect to
the scalar product
\bea
\langle h,g\rangle&=& \int dxdy\,
w(x)w(y)F(x-y)h(y)g(x)\ ,\\
F(x)&=&  \e^{\frac{x^2(1-a^2)}{8a^2}}
\left[\erf\Big(\frac{x\sqrt{1-a^2}}{\sqrt{8a^2}}
+\frac{m}{\sqrt{2a^2(1-a^2)}}\Big)
+
\erf\Big(\frac{x\sqrt{1-a^2}}{\sqrt{8a^2}}-\frac{m}{\sqrt{2a^2(1-a^2)}}\Big)
\right].
\nn
\eea
Because the polynomials obviously have parity, $R_j(-x)=(-)^jR_j(x)$,
and because the weight is antisymmetric, $F(-x)=-F(x)$, it holds
\be
\langle R_{2j}, R_{2l}\rangle=0=\langle R_{2j+1}, R_{2l+1}\rangle\ .
\ee
Furthermore, the even polynomials are skew-orthogonal to all
polynomials of lower degree, 
as can be seen from eq. (\ref{preskew}),
\be
\langle R_{2j}, R_{2l+1}\rangle\ =\ 0\ ,\ \mbox{for}\ \ j>l.
\label{skewjl}
\ee
It remains to investigate the remaining cases. For $j<l$ we have
\bea
\langle R_{2j}, R_{2l+1}\rangle&=& 
-2a^2 \int dxdy w(y) F(x-y) R_{2j}(y) \frac{d}{dx}\Big[
w(x) R_{2l}(x)\Big]\nn\\
&=&-2a^2 \int dxdy w(x)w(y) \frac{dF(x-y)}{dy} R_{2j}(y) R_{2l}(x)\nn\\
&=&2a^2 \int dxdy w(x) F(x-y) R_{2l}(x) \frac{d}{dy}\Big[
w(y) R_{2j}(y)\Big]=0\ .
\eea
After a first integration by parts with respect to $x$ we have used
$\frac{dF(x-y)}{dx} =-\frac{dF(x-y)}{dy}$, and a second integration by
parts with respect to $y$ yields zero after using eq. (\ref{skewjl}).

It remains to compute the only nonvanishing product for $j=l$:
\bea
\langle R_{2j}, R_{2j+1}\rangle&=& 2a^2 \int dxdy w(x) w(y)
\frac{dF(x-y)}{dx} R_{2j}(y) R_{2l}(x)\nn\\
&=& \int dxdy w(x) w(y) \frac{1-a^2}{2}(x-y)F(x-y) R_{2j}(y) R_{2l}(x)\nn\\
&&+\sqrt{\frac{2a^2(1-a^2)}{\pi}}\e^{-\frac{m^2}{2a^2(1-a^2)}}
\int dxdy w(x) w(y)\left(
  \e^{\frac{m(x-y)}{2a^2}}+e^{\frac{-m(x-y)}{2a^2}}\right)R_{2j}(y)
R_{2l}(x)\nn\\
&=& a^2(1-a^2)\int dxdy \Big(\frac{dw(x)}{dx}
w(y)-w(x)\frac{dw(y)}{dy}\Big)F(x-y)R_{2j}(x)R_{2j}(y)\nn\\
&&+2\sqrt{\frac{2a^2(1-a^2)}{\pi}}\e^{-\frac{m^2}{2(1-a^2)}}
\int dx w(x-m) R_{2j}(x) \int dy w(y+m)R_{2j}(y) \ .
\eea
In the first step we simply integrate by parts. We then observe that
the first line in the last equation is proportional to $\langle
R_{2j}, R_{2j+1}\rangle$ because when the derivatives also act on the
polynomials $R_{2j}(x)$ and $R_{2j}(y)$ this gives zero as the
polynomials $R_{2j}$ are skew-orthogonal to all lower order
polynomials. Putting this term on the left hand side and using parity
we obtain
\be
a^2\langle R_{2j}, R_{2j+1}\rangle
\ =\ 2\sqrt{\frac{2a^2(1-a^2)}{\pi}}\e^{-\frac{m^2}{2(1-a^2)}}
\left(\int dx w(x-m) R_{2j}(x)\right)^2 \ .
\ee
In a last step we have to evaluate the integral over the weighted even
polynomials. Using eq. (\ref{Reresult}) we have
\bea
\int dx w(x-m)R_{2j}(x)&=& \frac{j!\,(2(1-a^2))^{j}}{(-)^j\sqrt{\pi}}
\int_{-\infty}^\infty  dx\
\e^{-\frac{(x-m)^2}{4a^2}}\int_{-\infty}^\infty  dt\ \e^{-t^2}
L_j\left( \frac{(x+2iat)^2-m^2}{2(1-a^2)}\right)\nn\\
&=& \frac{j!\,(2(1-a^2))^{j}}{(-)^j\sqrt{\pi}}
\frac{1}{2a}\int_{-\infty}^\infty  dxds\ \e^{-\frac{s^2+x^2}{4a^2}}
L_j\left( \frac{x^2+2isx-s^2+2m(x+is)}{2(1-a^2)}\right)\nn\\
&=&\frac{j!\,(2(1-a^2))^{j}}{(-)^j\sqrt{\pi}}2a\pi\ .
\label{normint}
\eea
In the last but one line we can change to complex coordinates $z=x+is$
and to an integration over the full complex plane. Because of the
argument of $L_n(z(z+2m)/2(1-a^2))$ the angular integration will give
zero for all powers of the argument $z$, except for the constant term
$z^0$ which
is unity, $L_n(0)=1$. Collecting all factors we thus obtain
\be
r_j\ =\ \langle R_{2j}, R_{2j+1}\rangle\ =\
8\sqrt{2\pi}\ a\ 2^{2j} (j!)^2 (1-a^2)^{2j+\frac12}\e^{-\frac{m^2}{2(1-a^2)}}\ .
\ee

\end{appendix}


\end{document}